\documentclass{article}

\usepackage[dandb, final]{neurips_2025}

\usepackage[utf8]{inputenc} 
\usepackage[T1]{fontenc}    
\usepackage{hyperref}       
\usepackage{url}            
\usepackage{booktabs}       
\usepackage{amsfonts}       
\usepackage{nicefrac}       
\usepackage{microtype}      
\usepackage{xcolor}         
\usepackage{pifont}
\usepackage{multirow}
\usepackage{arydshln}
\usepackage{array}
\usepackage{hyperref}
\usepackage{graphicx}
\usepackage{subcaption}
\usepackage{makecell}
\usepackage{float}
\usepackage{verbatim}
\usepackage{epstopdf}
\title{CSI-Bench: A Large-Scale In-the-Wild Dataset for Multi-task WiFi Sensing}
\renewcommand{\arraystretch}{1.2}

\author{%
  Guozhen~Zhu, Yuqian~Hu, Weihang~Gao, Wei-Hsiang~Wang, Beibei~Wang, K.~J.~Ray~Liu \\
  Origin Research\\
  \texttt{\{guozhen.zhu,yuqian.hu,weihang.gao,weihsiang.wang,beibei.wang,ray.liu\}}\\
  \texttt{@originwirelessai.com } \\
}

\begin{document}

\maketitle

\begin{abstract}
WiFi sensing has emerged as a compelling contactless modality for human activity monitoring by capturing fine-grained variations in Channel State Information (CSI). Its ability to operate continuously and non-intrusively while preserving user privacy makes it particularly suitable for health monitoring. However, existing WiFi sensing systems struggle to generalize in real-world settings, largely due to datasets collected in controlled environments with homogeneous hardware and fragmented, session-based recordings that fail to reflect continuous daily activity.

We present CSI-Bench, a large-scale, in-the-wild benchmark dataset collected using commercial WiFi edge devices across 26 diverse indoor environments with 35 real users. Spanning over 461 hours of effective data, CSI-Bench captures realistic signal variability under natural conditions. It includes task-specific datasets for fall detection, breathing monitoring, localization, and motion source recognition, as well as a co-labeled multitask dataset with joint annotations for user identity, activity, and proximity. To support the development of robust and generalizable models, CSI-Bench provides standardized evaluation splits and baseline results for both single-task and multi-task learning. CSI-Bench offers a foundation for scalable, privacy-preserving WiFi sensing systems in health and broader human-centric applications.
\textit{Links: }\href{https://www.kaggle.com/datasets/guozhenjennzhu/csi-bench}{\textcolor{blue}{\textit{CSI-Bench Dataset}}};
\href{https://github.com/Jenny-Zhu/CSI-Bench-Real-WiFi-Sensing-Benchmark.git}{\textcolor{blue}{\textit{CSI-Bench Code}}}; 
\href{https://ai-iot-sensing.github.io/projects/project.html}{\textcolor{blue}{\textit{Project Page}}}

\end{abstract}

\section{Introduction}

Today’s smart IoT devices, such as smart speakers, smart bulbs, and various smart display devices, are commonly connected to home routers or mesh network hubs via WiFi. Beyond their primary role in communication, the WiFi signals between these devices inherently capture rich information about the surrounding environment through their propagation paths~\cite{liu_wang_2019,wificandomore2022, ray2024spmag}. This has positioned WiFi sensing as a compelling alternative to vision- or wearable-based systems for human monitoring in smart environments. By capturing fine-grained temporal and spatial variations in Channel State Information (CSI), commodity WiFi devices can infer a wide range of human-centric phenomena—from gross motor events such as falls to subtle physiological signals like breathing. These properties make WiFi sensing especially attractive for health-related applications in smart homes, where privacy, continuous operation, and ease of deployment are critical. Moreover, because these signals are already being transmitted by existing infrastructure, WiFi-based sensing enables non-intrusive, cost-effective, and passive monitoring without requiring additional sensors or user instrumentation.

Despite increasing research interest, existing WiFi sensing studies suffer from a fundamental limitation: a lack of large-scale, diverse, and real-world datasets. Most current datasets are collected in controlled laboratory settings, often using limited types of homogeneous hardware configurations and a narrow range of tasks. As a result, models trained on these datasets struggle to generalize to new users, devices, or environments, limiting their practical utility.

\begin{figure}[t]
    \centering
    \includegraphics[width=0.99\linewidth]{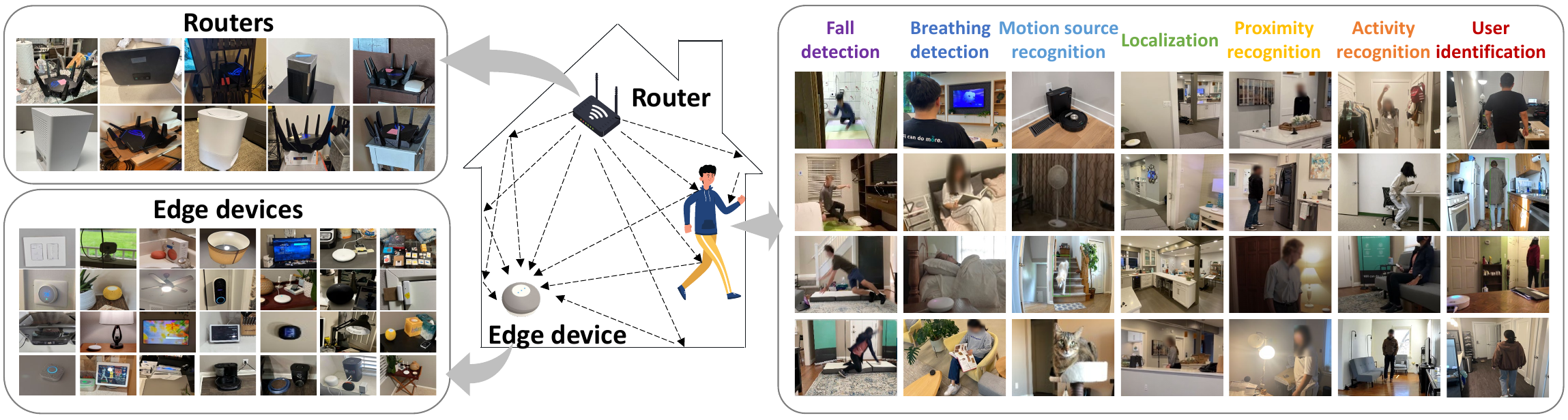}
    \caption{CSI-Bench overview. The benchmark features multiple commercial routers and IoT devices deployed in real homes and offices to collect CSI data. It supports a wide range of human-centric sensing tasks, enabling robust model development across diverse hardware setups and real-world scenarios.}
    \label{fig:benchmark-concept}
\end{figure}

To address these gaps, we introduce CSI-Bench, the first large-scale, in-the-wild benchmark dataset supporting multi-task WiFi sensing as illustrated in Figure \ref{fig:benchmark-concept}. Using commercial edge devices, CSI-Bench captures real-world signal variability across diverse environments, including apartments, multi-room houses, offices, and public indoor spaces. Data is recorded continuously from a broad spectrum of WiFi chipsets (Qualcomm, Broadcom, Espressif, MediaTek, and NXP), under both line-of-sight (LoS) and non-line-of-sight (NLoS) conditions, and during natural human activities with minimal intervention.

CSI-Bench advances the field in three key ways:

\noindent\textbf{Large-scale, real-world coverage.} The dataset spans over \textbf{461} hours of CSI data from \textbf{35} users, \textbf{26} distinct environments, and \textbf{16} device configurations. It reflects realistic deployment conditions with background interference, user mobility, and ambient network traffic.

\noindent\textbf{Multi-task and co-labeled annotations.} We provide both single-task specialist datasets (e.g., fall detection, breathing monitoring, localization, and motion source recognition) and a multi-task dataset with joint labels for user identification, activity recognition, and proximity estimation. The co-labeled samples enable efficient multi-task learning and low-latency inference on resource-constrained edge devices.

\noindent\textbf{Standardized benchmarking protocols.} We establish strong baselines under supervised learning and multi-task learning. Our findings highlight generalization gaps and the promise of parameter-efficient multi-task learning.

CSI-Bench aims to catalyze robust model development for passive, privacy-preserving WiFi sensing. By offering a unified platform for realistic, diverse, and reproducible evaluation, it provides a foundation for scalable AI applications in smart health, home monitoring, and beyond.

\section{Related Work}
\subsection{WiFi Sensing}
Compared to vision-, audio-, or wearable-based systems, WiFi sensing offers a scalable, privacy-preserving, and non-intrusive alternative or complementary solution for continuous monitoring in smart environments and healthcare applications~\cite{Wi-Fi2020, electronics12234858, SOTO202299,SrcSense}. WiFi sensing has demonstrated substantial potential in tasks such as activity recognition~\cite{bing2021that, Meneghello2023activity}, gesture detection~\cite{sai2023gesture, yang2019gesture}, indoor tracking and localization~\cite{wang2017localization, wang2024localization,EZMap,floorplan}, fall detection~\cite{FallAware}, proximity detection~\cite{proximity}, and vital sign monitoring~\cite{wang2024vital, gao2020vital}. However, most existing studies rely on data collected in constrained settings, which limits generalization to diverse users, hardware platforms, and real-world deployment scenarios.


\subsection{WiFi Sensing Dataset}


A number of WiFi sensing datasets have contributed valuable resources to the community. Widar3.0~\cite{widar3} offers large-scale CSI data for gesture recognition using Intel 5300 NICs~\cite{NIC5300}. SignFi~\cite{signfi} focuses on sign language recognition, capturing fine-grained hand gestures. MM-Fi~\cite{yang2023mm} enables cross-modal analysis by combining WiFi CSI with synchronized video and depth data. XRF55~\cite{wang2024xrf55} introduces a large corpus of RF-based activity data for action recognition. Additional datasets such as ARIL~\cite{aril}, CSIDA~\cite{csida} support tasks like activity recognition and localization.

While these datasets have advanced the field, they share several limitations as illustrated in Table \ref{tab:dataset compare}. First, most are confined to controlled laboratory settings, offering limited variability in user behavior, device types, and environmental complexity. Second, they primarily support single-task scenarios, lacking the multi-task supervision needed for training general-purpose models. Third, nearly all rely on the Intel 5300 chipset, which does not support continuous CSI recording. As a result, data is collected in fragmented, pre-scripted sessions using manual triggers, which limits dataset scale and fails to capture users’ natural daily activities.
There remains a growing demand for a unified benchmark that reflects the complexity of real-world deployments, supports multiple sensing tasks, and enables evaluation across diverse users, environments, and hardware platforms. 
To address this need, we introduce CSI-Bench, a large-scale in-the-wild benchmark for passive WiFi sensing.


\begin{table}[t]
\centering
\caption{Comparison of CSI-Bench with published datasets.}
\label{tab:dataset compare}
\resizebox{\textwidth}{!}{%
\begin{tabular}{lcccccccc}
    \toprule
    \textbf{Dataset (Year)} & \textbf{Platform} & \textbf{\#Edge Device Type} &\textbf{\#Samples} & \textbf{\#Tasks} & \textbf{\#Envs} & \textbf{\#Users} & \textbf{In-the-Wild}  \\
    \midrule
   

    \textbf{WiAR~\cite{wiar} (2019)} & Intel 5300 NIC   & 1 & 4.8k & 1 & 3 & 10 &  \ding{55} \\
    \textbf{ARIL\cite{aril} (2019)} & USRP  & 1  & 1.4k & 2 & 1 & 1 & \ding{55} \\
    \textbf{Widar3.0~\cite{widar3} (2021)} & Intel 5300 NIC & 1 & 271.1k & 1 & 3 & 16 & \ding{55} \\
    \textbf{XRF55~\cite{wang2024xrf55} (2024)} & Intel 5300 NIC & 1 & 42.9k & 1 & 4 & 39& \ding{55} \\
    \textbf{SignFi~\cite{signfi} (2018)} &  Intel 5300 NIC  & 1 & 14.3k & 1 & 2 & 5 & \ding{55} \\
    \textbf{WiMANs~\cite{huang2024wimans} (2024)} & Intel 5300 NIC & 1 & 11.3k  & 3 & 3 & 5 & \ding{55} \\
    \textbf{CSIDA~\cite{csida} (2021)} & Intel 5300 NIC& 1 & 3k & 1 & 2 & 5 & \ding{55} \\
    \textbf{MM-Fi~\cite{yang2023mm} (2023)} & Atheros CSI Tool & 1 & 1.1k & 1 & 4 & 40  & \ding{55} \\
    \midrule
     \multirow{3}{*}{\textbf{CSI-Bench}} & Broadcom  & \multirow{3}{*}{16} & \multirow{3}{*}{ 231.6k} & \multirow{3}{*}{7} & \multirow{3}{*}{26} & \multirow{3}{*}{35} & \multirow{3}{*}{\ding{51}} \\
    & Qualcomm, MediaTek & &  &  &  &  &  \\
    & Espressif, NXP    & &  &  &  &  &  \\
    \bottomrule
\end{tabular}%
}
\end{table}


\section{Dataset Collection}

\subsection{Overview}
\begin{figure}[t]
    \centering
    \includegraphics[width=0.99\linewidth]{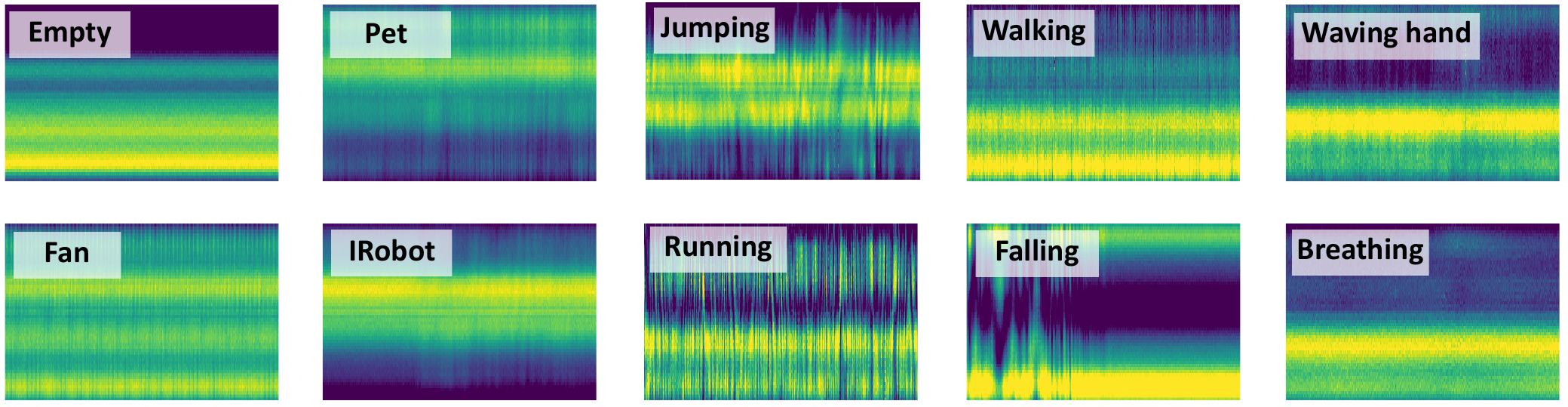}
    \caption{Representative CSI samples are shown for various scenarios, including human actions (jumping, running, walking, hand waving, falling, breathing), non-human motions (pet movement, iRobot, fan), and empty environments. In each sample, the x-axis represents time, and the y-axis represents the subcarrier index.}
    \label{fig:csi-example}
\end{figure}

To support robust and generalizable WiFi sensing research, we build a diverse collection of datasets captured in real-world environments using commercial WiFi devices. CSI-Bench spans over \textbf{461 hours} of CSI recordings across \textbf{35 unique users}, \textbf{26 environments}, and \textbf{16 device types}, covering both routers and edge devices operating under varied network conditions. Data is collected in homes, offices, and public indoor areas with minimal control over ambient interference or user behavior. Each dataset is designed to support one or more sensing tasks, including fall detection (Fall), breathing monitoring (Breath), localization (Loc.), human activity recognition (HAR), user identification (UID), and proximity estimation (Prox.). Representative CSI samples illustrating task-specific signal patterns are visualized in Figure~\ref{fig:csi-example}. The following section details the hardware, environments, and collection protocols used to capture the datasets.

\subsection{Devices and Hardware Setup} 
\textbf{Hardware.} 
To emulate the heterogeneity of real-world WiFi sensing deployments, we employ a diverse set of commercial WiFi routers and edge IoT devices commonly found in residential and commercial environments. The selected devices span five major chipset vendors, including Qualcomm, MediaTek, Broadcom, Espressif, and NXP~\cite{Qualcomm,NXP,Broadcom,Espressif}, and cover a broad spectrum of hardware configurations, including 1×1 to 2×2 MIMO and 1×4 antenna setups. All devices support IEEE 802.11 n/ac/ax standards, operating across both 2.4 GHz and 5 GHz bands with channel bandwidths of 20, 40, and 80 MHz.
These heterogeneous devices are intentionally chosen to reflect the real IoT ecosystem deployed in homes, offices, and small businesses, where heterogeneous devices with varying wireless capabilities coexist in complex indoor environments. Their detailed specifications, including chipset model, antenna configuration, bandwidth, frequency band, and empirically measured average RSSI, are listed in Appendix~\ref{app:dataset} (Table~\ref{tab:device_summary}).

\textbf{CSI extraction and synchronization.}
In our system, IoT client devices periodically transmit CSI packets to routers at two sounding rates: 100 Hz for general sensing tasks and 30 Hz for breathing detection, accommodating different temporal dynamics. Given the distributed nature of these devices, propagation delays and clock drifts cause misalignment in CSI data streams. To address this, the router coordinates data collection by sending batch requests with defined time windows, asking devices to record and upload CSI within the same interval. Each device uses its own system clock to timestamp the data, which allows us to later align the streams in software. Routers handle CSI extraction, buffering, and data upload to cloud servers, running either Linux or FreeRTOS depending on their chipset.

\textbf{CSI format.} 
Due to hardware diversity, the CSI data in CSI-Bench varies in subcarrier granularity, antenna configurations, and supported bandwidths across different chipset architectures. For example, the NXP 88W8997 provides a 2×2 MIMO configuration with 58 subcarriers at 40 MHz on 5 GHz, while the ESP32-S3, with a 1×1 setup, captures 64 subcarriers at 20 MHz on 2.4 GHz. Qualcomm IPQ4019/IPQ4018 devices offer a 1×2 MIMO configuration, supporting 128 subcarriers at 40 MHz and 256 subcarriers at 80 MHz on 5 GHz. In contrast, the Broadcom BCM4345 employs a 1×4 antenna configuration, providing only 14/28 subcarriers at 20/40 MHz due to proprietary subcarrier grouping. These variations ensure CSI-Bench captures a wide spectrum of signal characteristics, enabling comprehensive evaluation of model generalization across heterogeneous hardware platforms.



\subsection{Continuous Data Recording}
\label{subsec:continuous_data_recording}
To overcome the limitations of prior works that typically rely on controlled environments or predefined protocols, we develop an integrated pipeline enabling scalable, in-the-wild CSI data collection across diverse residential settings. Leveraging commercial routers with developer-accessible CSI extraction, cloud infrastructure, and user-friendly annotation tools, our system unobtrusively captures large-scale CSI data from everyday WiFi usage without device-side modifications.

We collaborate with multiple router chipset vendors, who provided firmware and drivers with CSI extraction capabilities enabled, along with proprietary CSI capture utilities for CSI extraction. Building on this, we develop our own tools to programmatically capture and manage CSI data. Specifically, we design separate tools for Linux or FreeRTOS~\cite{FreeRTOS}, each design to send commands from the Linux application layer directly to the WLAN kernel module, enabling continuous collection and buffering of CSI from all registered devices into unified binary files, which are periodically uploaded to cloud storage via AWS S3 APIs~\cite{AWSS3API}. Each file is timestamped using the router's local system time embedded in the filename, ensuring straightforward temporal alignment across deployments. Upload frequency dynamically adjusts based on device count and bandwidth utilization.

We also develop a lightweight user annotation tool integrated into Google Spreadsheet~\cite{GoogleSheets}, allowing users to optionally log daily activities—such as waking up, sleeping, leaving or returning home, room occupancy, or inactivity—by tapping buttons that record local timestamps. This design minimizes user effort while ensuring accurate temporal alignment between activity logs and CSI data. An illustration of the annotation tool is provided in Appendix \ref{app:annotation_tool}. Our system queries and retrieves CSI files matching these events, concatenates the relevant segments, and refines alignment using embedded packet-level timestamps, resulting in precisely labeled CSI data segments. We collect CSI of motion from non-human sources like pets and cleaning robot when users are not home. When possible, time-aligned external information is collected through camera recordings and local sensor logs to annotate non-human motions or highlight environmental changes.

This pipeline enables extensive, accurately labeled CSI data collection reflective of authentic user behaviors and diverse environments, supporting a wide range of large-scale research applications.

\subsection{Environments and Contexts} 
We collect our data across a broad range of environments, including compact apartments, multi-room houses, offices, hallways, and open indoor public spaces, as detailed in Appendix~\ref{app:environments}. These settings introduce diverse physical characteristics, including complex layouts, clutter, variable wall materials, and occlusions, that significantly affect signal propagation.

Unlike prior datasets collected under controlled conditions, our data captures CSI under authentic, in-the-wild conditions. Devices were positioned freely by users, and data was recorded continuously during natural daily activities. Consequently, the CSI reflects realistic variability introduced by NLoS links, neighboring motion, background activity from appliances, WiFi traffic, and environmental factors such as wind and even rain drops. This level of interference is critical for benchmarking the robustness of WiFi sensing models, particularly for healthcare applications where reliable and through-the-wall monitoring in uncontrolled home environments is essential.

\subsection{Data Collection Protocols} 
\label{sec:datacollectionprotocol}
Although participants are free to move naturally and perform tasks as they would in daily life, we implement basic data collection protocols to ensure consistency and repeatability. Each session begin with a brief calibration phase to verify device connectivity, synchronize timestamps, and confirm stable CSI logging. The recorded activities spans a range of motion patterns, including sitting still, walking, waving hands, and running through hallways. All participants signed a consent form prior to participation, with expenses around \$20 /hr. Data from non-human motion sources—such as pets, cleaning robots, and electrical appliances like fans—are collected when users are not present. Detailed task-specific data collection procedures are provided in Appendix~\ref{app:dataset}.

\subsection{Dataset Statistics}
\label{sec:data-stats}

CSI-Bench spans seven classification tasks with varied sensing objectives. Table~\ref{tab:task-summary} summarizes dataset scale and coverage, including the number of samples, recording duration, users, environments, and device types. This diversity reflects real-world deployment conditions and supports robust generalization benchmarking.

\begin{table}[t]
\centering
\scriptsize
\caption{Summary of tasks, dataset statistics, partitions, and evaluation protocols. \textit{ST} = single-task specialist, \textit{MT} = multi-task joint. }
\label{tab:task-summary}
\setlength{\tabcolsep}{5pt}
\begin{tabular}{lccccccc}
\toprule
\textbf{Task} & \textbf{\#Classes} & \textbf{Dataset} & \textbf{\#Samples} & \textbf{\#Users} & \textbf{\#Envs} & \textbf{\#Devices} & \textbf{Split, Setting} \\
\midrule
Fall Detection    & 2   & ST & 6.7k & 17 & 6 & 2 & 70/15/15, \textit{easy/med/hard} \\
Breath Detection  & 2   & ST & 100k & 3 & 3 & 6 & 70/15/15, \textit{easy/med/hard} \\
Motion Source Recognition      & 4   & ST & 60.9k & 35 & 10 & 1 & 70/15/15, \textit{easy/med/hard} \\
Room-level Localization     & 6   & ST & 7.1k & 8 & 6 & 8 & 70/15/15, \textit{easy/med/hard} \\
Proximity Recognition    & 4   & MT & 20.3k & 6 & 6 & 11 & 70/15/15, \textit{user/env/device} \\
Human Activity Recognition      & 5   & MT & 41.5k & 6 & 6 & 11 & 70/15/15, \textit{user/env/device} \\
User Identification     & 6 & MT & 20.3k & 6 & 6 & 11 & 70/15/15, \textit{device} \\
\bottomrule
\end{tabular}
\end{table}

\section{Data Quality and Preprocessing}

\subsection{CSI Quality Verification}

\begin{figure}[t]
    \centering
    \begin{subfigure}[b]{0.6\linewidth}
        \centering
        \includegraphics[width=\linewidth]{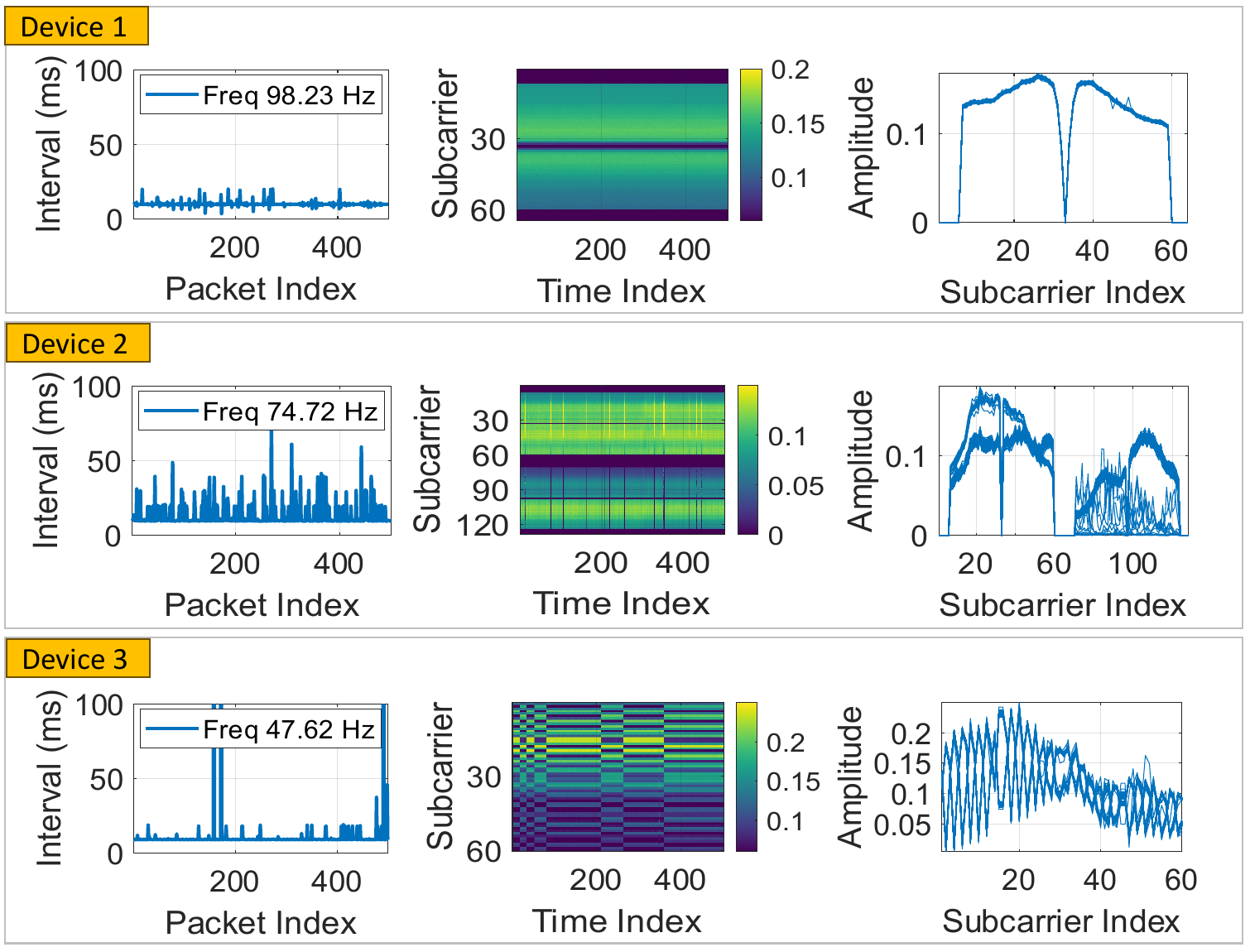}
        \caption{}
        \label{fig:csi_quality_check}
    \end{subfigure}
    \hfill
    \begin{subfigure}[b]{0.38\linewidth}
        \centering
        \includegraphics[width=\linewidth]{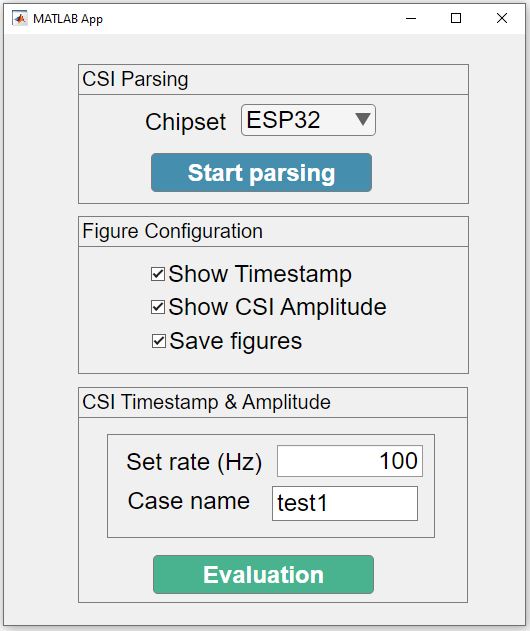}
        \caption{}
        \label{fig:matlab_tool}
    \end{subfigure}
    \caption{ MATLAB-based CSI verification tool. (a) Visualization of CSI quality from three devices, showing variations in sampling interval, time-subcarrier heatmap, and amplitude response. (b) User interface for parsing and evaluating CSI data, supporting timestamp checks, amplitude analysis, and figure export to ensure data reliability in CSI-Bench.}
    \label{fig:matlab_tool_combined}
\end{figure}

\textbf{Motivation.}  
CSI quality checking is critical for ensuring data reliability, as raw measurements often suffer from signal dropouts, high noise levels, or inconsistent timestamps. These issues can arise due to differences in chipset design, CSI extraction algorithms, hardware configurations (e.g., antenna layout, RF circuitry), and deployment conditions. As illustrated in Figure~\ref{fig:csi_quality_check}, the CSI quality varies from device to device. Device 1 exhibits the best CSI quality, with consistent temporal patterns and a stable sampling rate near the nominal 30 and 100 Hz. Device 2 shows moderate quality with occasional outliers and a lower sampling rate, while Device 3 suffers from the poorest quality, marked by irregular sampling intervals and temporal clustering of CSI frames. Given the diverse hardware platforms and settings in CSI-Bench, these quality variations must be systematically addressed to enable meaningful benchmarking. 

\textbf{Verification tool.} To systematically assess and ensure CSI data quality, we adopt a structured evaluation framework introduced in an existing work~\cite{yuqian2024csiveri}, which models CSI verification as a multilayered pipeline. Each layer of this pipeline targets a specific aspect of data integrity using customized metrics, covering timestamp consistency, CSI amplitude stability, and other modality-specific characteristics. This design allows us to characterize various perspectives of CSI quality and adapt the evaluation to different sensing tasks. In the context of CSI-Bench, we apply this framework to filter out samples with timestamp irregularities, unstable or flat CSI amplitude, and signal dropout, ensuring that only reliable traces are included in the benchmark. The CSI verification tool is implemented in MATLAB, as shown in Figure~\ref{fig:matlab_tool}, to facilitate systematic quality control before incorporating data into CSI-Bench.


\subsection{CSI Preprocessing Pipeline}
\label{subsec:csi preprocessing pipeline}
\textbf{Amplitude extraction.}
In real-world measurements, CSI is often corrupted by phase noise caused by timing and frequency synchronization offsets, as well as additive thermal noise. In the literature, two main approaches are used to handle phase distortions: phase cleaning~\cite{cc2017phase,kun2017widar,qinyi2017} and phase elimination~\cite{wei2015,wang2017wifall,feng2019widetect}. Phase cleaning aims to correct the distorted phase but cannot fully eliminate initial phase offsets, making it less reliable for consistent processing across diverse devices. Therefore, in our benchmark, we adopt the phase elimination approach. Specifically, if the extracted CSI at time $t$ and subcarrier frequency $f$ is represented as $H(f,t)$, we use the amplitude $|H(f,t)|$ as input, eliminating the unreliable phase component.

\textbf{Data segmentation.}
To facilitate supervised model training, the collected CSI data is segmented into fixed-duration, non-overlapping samples. For tasks including Fall Detection, Localization, Motion Source Recognition, and the Multi-Task dataset, we segment CSI data into 5-second intervals. For the Breathing Detection dataset, considering the slower temporal variations inherent to respiration signals, we segment the CSI data into 10-second intervals.
We also provide the unsegmented CSI recordings of part of our dataset to support


\textbf{Amplitude normalization.}
To mitigate the effects of varying signal strengths, we normalize each CSI sample across all subcarriers and time indices by removing the mean and scaling by the standard deviation. 
This ensures consistent scaling across samples while preserving the relative temporal–spectral dynamics within each sample. 
The normalized CSI is computed as $\hat{H}(f_k, t) = \frac{H(f_k, t) - \mu_H}{\sigma_H + \epsilon}$
where $\mu_H$ and $\sigma_H$ denote the mean and standard deviation of $H(f_k, t)$ over the entire CSI matrix, and $\epsilon$ is a small constant added for numerical stability.

\textbf{Subcarrier standardization.}
Due to hardware differences, the number of subcarriers in CSI samples can vary across different platforms, leading to inconsistent input shapes along the frequency dimension. To standardize the data, we select a fixed number of subcarriers and apply zero-padding or clipping in the frequency dimension as needed. This ensures all samples have consistent input shapes across the dataset.

\section{Benchmark Design}
\label{sec:benchmark}

\subsection{Task Suite and Metrics}
\label{sec:tasks}

CSI-Bench supports a suite of supervised classification tasks for WiFi sensing, covering key applications in health monitoring and ambient intelligence. Each task operates on a fixed-length CSI tensor $\mathbf{X} \in \mathbb{R}^{C \times K \times T}$, where $C$ is the channel count, $K$ is the standardized subcarrier dimension over antenna arrays, and $T$ is the temporal length of samples (5 seconds for most tasks, and 10 seconds for breathing detection).

\noindent\textbf{Single-task specialized dataset.} The benchmark includes four single-task datasets: \textit{Fall Detection} (binary classification of fall vs. non-fall), \textit{Breathing Detection} (binary detection during sleep, sampled at 30 Hz), \textit{Motion Source Recognition} (four-class classification of human, pet, robot, and fan motion), and \textit{Room-Level Localization} (six-way classification of the user location). These are evaluated independently using dedicated datasets.

\noindent\textbf{Multi-task joint dataset.} A multi-task dataset contains co-labeled samples for three tasks: \textit{Human Activity Recognition} (five-class classification), \textit{User Identification} (multi-class over 6 users), and \textit{Proximity Recognition} (four-class distance estimation). This enables parameter-efficient multi-task training with a shared backbone and task-specific heads.



All tasks are evaluated using overall accuracy and weighted F1-score. Accuracy provides a global measure of classification correctness, while the weighted F1-score accounts for class imbalance by averaging per-class F1-scores weighted by class frequency. This is especially relevant for tasks with skewed distributions such as fall detection or proximity recognition.

\subsection{Evaluation Protocols}
\label{sec:eval-protocols}

CSI-Bench provides standardized train/validation/test splits for all tasks to ensure fair comparison and reproducibility. For each dataset, 70\% of samples are used for training, 15\% for validation, and the remaining 15\% for testing, with class balance and environment distribution preserved. Evaluation protocols and statistics for each task are summarized in Table~\ref{tab:task-summary}.

To evaluate real-world robustness, each test sample is annotated with a difficulty level—Easy, Medium, or Hard—based on signal quality, environment, and subject complexity. For the multi-task dataset, we define three out-of-distribution (OOD) splits—cross-user, cross-environment, and cross-device—reflecting domain shifts in deployment. These settings enable systematic robustness and generalization evaluation. Full details are provided in Appendix~\ref{app:dataset}.

\medskip


\subsection{Baseline Models}
\label{sec:baselines}

To establish reference performance and benchmark learning effectiveness on CSI-Bench, we implement a suite of baseline models across single-task supervised and multi-task learning settings.

\noindent\textbf{Supervised learning.} We evaluate representative architectures spanning fully connected networks (MLP)~\cite{MLP}, recurrent models (LSTM)~\cite{LSTM1997}, convolutional backbones (ResNet-18)~\cite{kaiming2016}, and transformer-based sequence learners, including Vision Transformer (ViT)~\cite{ViT2021}, PatchTST~\cite{PatchTST2023}, and TimeSformer-1D~\cite{TimeSformer2021}. All models are trained independently on each task using the corresponding specialist dataset. Input CSI tensors are amplitude-only with hyperparameters tuned using validation performance.

\noindent\textbf{Multi-task learning.} To explore parameter efficiency and cross-task knowledge sharing, we also implement multi-task learning using a shared backbone with lightweight task-specific adapters~\cite{Caruana1997}. We adopt the same backbones as in the supervised setting and attach low-rank (LoRA) adapters~\cite{LoRA2021} and separate classification heads for each task. During training, task-labeled samples are drawn from the joint multi-task dataset, and optimization proceeds with shared backbone updates and task-specific losses.

All models are trained using the AdamW optimizer~\cite{Adam19} with a cosine learning rate schedule and early stopping. Detailed architecture configurations and training hyperparameters are provided in Appendix~\ref{app:model-details}.

\begin{table}[t]
\centering
\caption{Performance comparison of supervised models across four core WiFi sensing tasks. Accuracy (Acc) and F1-score are reported as mean ± std (\%) over three runs.}
\label{tab:supervised-results}
\resizebox{\textwidth}{!}{%
\begin{tabular}{l
cc cc cc cc}
\toprule
\multirow{2}{*}{\textbf{Model}} 
& \multicolumn{2}{c}{\textbf{Fall Detection}} 
& \multicolumn{2}{c}{\textbf{Breathing Detection}} 
& \multicolumn{2}{c}{\textbf{Room-Level Localization}} 
& \multicolumn{2}{c}{\textbf{Motion Source Recognition}} \\
\cmidrule(lr){2-3} \cmidrule(lr){4-5} \cmidrule(lr){6-7} \cmidrule(lr){8-9}
 & Acc & F1 & Acc & F1 & Acc & F1 & Acc & F1 \\
\midrule
MLP   ~\cite{MLP}       & 92.16 \textpm {\scriptsize 0.91} & 92.17 \textpm {\scriptsize 0.92} & 97.59 \textpm {\scriptsize 0.08} & 97.59 \textpm {\scriptsize 0.08} & 87.14 \textpm {\scriptsize 0.80} & 86.90 \textpm {\scriptsize 0.83} & 98.86 \textpm {\scriptsize 0.07} & 98.86 \textpm {\scriptsize 0.07} \\
ResNet-18 ~\cite{kaiming2016}   & 94.88 \textpm {\scriptsize 0.26} & 94.89 \textpm {\scriptsize 0.26} & 98.58 \textpm {\scriptsize 0.17} & 98.58 \textpm {\scriptsize 0.17} & 100.00 \textpm {\scriptsize 0.00} & 100.00 \textpm {\scriptsize 0.00} & 99.56 \textpm {\scriptsize 0.07 } & 99.56 \textpm {\scriptsize 0.07} \\
LSTM~\cite{LSTM1997} & 94.93 \textpm {\scriptsize 0.51} & 94.92 \textpm {\scriptsize 0.50} & 98.62 \textpm {\scriptsize 0.17} & 98.62 \textpm {\scriptsize 0.17} & 99.12 \textpm {\scriptsize 0.27} & 99.12 \textpm {\scriptsize 0.26} & 98.42 \textpm {\scriptsize 0.19} & 98.42 \textpm {\scriptsize 0.19} \\
Transformer~\cite{attention2017}  & 94.28 \textpm {\scriptsize 0.72} & 94.26 \textpm {\scriptsize 0.72} & 98.64 \textpm {\scriptsize 0.19} & 98.64 \textpm {\scriptsize 0.19} & 99.27 \textpm {\scriptsize 0.22} & 99.27 \textpm {\scriptsize 0.22} & 98.61 \textpm {\scriptsize 0.27} & 98.61 \textpm {\scriptsize 0.27} \\
ViT ~\cite{ViT2021}       & 93.58 \textpm {\scriptsize 0.71} & 93.59 \textpm {\scriptsize 0.70} & 98.63 \textpm {\scriptsize 0.17} & 98.63 \textpm {\scriptsize 0.17} & 99.94 \textpm {\scriptsize 0.11} & 99.94 \textpm {\scriptsize 0.11} & 98.74 \textpm {\scriptsize 0.10} & 98.74 \textpm {\scriptsize 0.10} \\
PatchTST~\cite{PatchTST2023}     & 94.03 \textpm {\scriptsize 0.74} & 94.03 \textpm {\scriptsize 0.73} & 98.84 \textpm {\scriptsize 0.13} & 98.84 \textpm {\scriptsize 0.13} & 99.91 \textpm {\scriptsize 0.10} & 99.91 \textpm {\scriptsize 0.10} & 98.86 \textpm {\scriptsize 0.19} & 98.86 \textpm {\scriptsize 0.19} \\
TimeSformer-1D~\cite{TimeSformer2021} & 93.86 \textpm {\scriptsize 1.16} & 93.87 \textpm {\scriptsize 1.13} & 98.68 \textpm {\scriptsize 0.21} & 98.68 \textpm {\scriptsize 0.21} & 100.00 \textpm {\scriptsize 0.00} & 100.00 \textpm {\scriptsize 0.00} & 98.38 \textpm {\scriptsize 0.17} & 98.39 \textpm {\scriptsize 0.17} \\
\bottomrule
\end{tabular}
}
\end{table}

\begin{table}[t]
\centering
\caption{Comparison of task-specific and multi-task training for the Transformer model across shared-data tasks. The improvements ($\Delta$) are reported as mean ± std (\%) over three runs.}
\label{tab:multitask_results}
\scriptsize
\begin{tabular}{l
cc cc cc}
\toprule
\multirow{2}{*}{\textbf{Task}} 
& \multicolumn{2}{c}{\textbf{Task-Specific Training}} 
& \multicolumn{2}{c}{\textbf{Multi-Task Joint Training}} 
& \multicolumn{2}{c}{\textbf{Improvement}} \\
\cmidrule(lr){2-3} \cmidrule(lr){4-5} \cmidrule(lr){6-7}
 & Acc & F1 & Acc & F1 & $\Delta$Acc & $\Delta$F1 \\
\midrule
Human Activity Recognition 
& 75.40 \textpm {\scriptsize 0.93} & 75.49 \textpm {\scriptsize 0.73} 
& 88.06 \textpm {\scriptsize 0.76} & 86.00 \textpm {\scriptsize 2.05} 
& +\textbf{12.66} & +\textbf{10.51} \\
User Identification        
& 99.51 \textpm {\scriptsize 0.32} & 99.51 \textpm {\scriptsize 0.32} 
& 99.55 \textpm {\scriptsize 0.06} & 99.70 \textpm {\scriptsize 0.27} 
& +0.04 & +0.19 \\
Proximity Recognition      
& 77.52 \textpm {\scriptsize 3.13} & 77.35 \textpm {\scriptsize 3.24} 
& 86.41 \textpm {\scriptsize 1.97} & 87.09 \textpm {\scriptsize 1.46} 
& +\textbf{8.89} & +\textbf{9.74} \\
\bottomrule
\end{tabular}
\end{table}

\subsection{Baseline Evaluation}
\label{sec:results}

We report performance on all tasks using both standard supervised learning baselines. Table~\ref{tab:supervised-results} summarizes accuracy and weighted F1-score for supervised models trained on the specialist datasets. Among the models, transformer-based architectures—particularly TimeSformer-1D and PatchTST—consistently achieve strong performance, highlighting their effectiveness in capturing temporal dynamics in high-dimensional CSI data. Simpler models such as MLP and LSTM perform adequately on some tasks but show clear limitations in harder cases.

Multi-task learning results are presented in Table~\ref{tab:multitask_results}. Compared to task-specific training, our multi-task models with a shared Transformer backbone and lightweight adapter-based heads achieve improved performance across multiple tasks. These findings highlight the effectiveness of joint training in capturing shared representations while preserving task-specific specialization through adapters. They also suggest that multi-task learning can improve generalization in real-world settings where sensing tasks are naturally co-located and co-labeled.

In addition to strong performance, our multi-task framework significantly reduces model complexity and training cost. By consolidating three single-task Transformers into a single backbone with task-specific adapters, we reduce the total parameter count by over 60\%. This compression is achieved without degrading task performance. Moreover, because all tasks are trained jointly in a single pass, the wall-clock training time is reduced by nearly $3\times$ compared to training separate models for each task. These gains in model size and training efficiency make our approach especially suitable for deployment on resource-constrained edge devices, where memory and compute budgets are limited.

We also report task-wise performance stratified by difficulty levels (Easy, Medium, Hard) for the single-task datasets in Appendix~\ref{app:difficulty-results}. Performance drops on hard samples for tasks like fall detection due to signal degradation, cluttered environments, and hardware diversity, reinforcing the need for deployment-aware evaluation.


\subsection{Evaluation on OOD Splits}
\label{app:ood-results}
We evaluate the Transformer-based multi-task model under three OOD conditions: cross-device, cross-environment, and cross-user (Table \ref{tab:ood-table}). 
Compared with the in-distribution results in Table \ref{tab:multitask_results}, all OOD accuracies and F1-scores drop substantially, indicating that models trained on seen domains fail to generalize effectively to unseen users, environments, or hardware. This degradation reflects WiFi CSI's strong sensitivity to device, environmental, and user variations, revealing a significant generalization gap between in-distribution and OOD domains. The results highlight the need for domain-adaptive, calibration-free learning frameworks to improve real-world robustness in WiFi sensing.
More detailed results and analysis can be found in Appendix \ref{app:ood-results}.

\begin{table}[t]
\centering
\caption{Cross-domain performance of Transformer model on three tasks. Accuracy (Acc) and F1-score are reported as mean ± std (\%) over three runs.}
\label{tab:ood-table}
\scriptsize
\begin{tabular}{l cc cc cc}
\toprule
\multirow{2}{*}{\textbf{Task}} & \multicolumn{2}{c}{\textbf{Cross-Device}} & \multicolumn{2}{c}{\textbf{Cross-Env}} & \multicolumn{2}{c}{\textbf{Cross-User}} \\
\cmidrule(lr){2-3}\cmidrule(lr){4-5}\cmidrule(lr){6-7}
&Acc & F1 & Acc & F1 & Acc & F1 \\
\midrule
Human Activity Recognition & 61.82 \textpm {\scriptsize 0.95} & 57.80 \textpm {\scriptsize 0.78} & 54.92 \textpm {\scriptsize 0.98} & 47.17 \textpm {\scriptsize 1.12} & 54.72 \textpm {\scriptsize 0.84} & 46.67 \textpm {\scriptsize 1.00} \\
Human Identification & 59.94 \textpm {\scriptsize 0.77} & 59.81 \textpm {\scriptsize 0.96} & / & / & / & / \\
Proximity Recognition & 30.68 \textpm {\scriptsize 3.11} & 28.76 \textpm {\scriptsize 3.51} & 29.67 \textpm {\scriptsize 1.63} & 27.12 \textpm {\scriptsize 1.79} & 30.26 \textpm {\scriptsize 1.94} & 25.97 \textpm {\scriptsize 2.26} \\
\bottomrule
\end{tabular}
\vspace{1ex}
\end{table}

\subsection{Discussion and Takeaways}
\label{sec:discussion}

CSI-Bench enables scalable research on high-dimensional CSI-based sensing under real-world conditions. Its large scale, diverse hardware coverage, and co-labeled tasks support the development of unified multi-task models for on-device health monitoring. Multi-task learning yields competitive performance while significantly reducing model size and inference cost, making it well-suited for resource-constrained edge deployment.
However, performance drops notably under OOD settings, particularly in cross-device scenarios, exposing persistent generalization challenges. Failure cases often arise from hardware heterogeneity, cluttered environments, or degraded signal quality.
Overall, CSI-Bench offers a realistic and comprehensive testbed for developing robust, efficient, and generalizable WiFi sensing systems in unconstrained environments.

\section{Limitations}
\label{sec:limitation}
The dataset uses amplitude-only CSI features due to phase instability across platforms. While this is practical, it limits exploration of techniques that exploit calibrated phase or angle-of-arrival information. CSI-Bench is designed around classification tasks. Extensions to regression (e.g., continuous  sign estimation) and more temporally structured tasks (e.g., long-term activity tracking) are promising but not yet included. We release all data, tools, and splits to support community-driven extensions and improvements.

\section{Conclusion}
\label{sec:conclusion}

We introduce CSI-Bench, a large-scale, in-the-wild benchmark dataset designed to advance research in WiFi-based sensing for health and human-centric applications. Collected using commercial WiFi edge devices deployed in real residential settings, CSI-Bench captures natural signal variability across users, devices, and environments—providing a realistic foundation for developing deployable, privacy-preserving WiFi sensing systems.
The dataset includes single-task datasets for fall detection, breathing monitoring, localization, and motion source recognition, as well as a co-labeled multi-task dataset supporting user identification, activity recognition and proximity recognition. This enables the development of multi-task models that support efficient joint inference while allowing rigorous evaluation under both in-distribution and out-of-distribution conditions.

To the best of our knowledge, CSI-Bench is the largest available dataset of its kind and can enable learning pipelines that benefit from high-dimensional CSI signals, diverse commercial edge devices, and real-world data (“in the wild”). Beyond the dataset, CSI-Bench includes a suite of baseline models and training protocols under supervised and multi-task settings. Our results show that multi-task learning can reduce model size and inference cost while maintaining competitive accuracy, making it suitable for health monitoring on resource-constrained devices.
At the same time, performance drops under domain shifts highlight the need for future research on adaptive and generalizable sensing models. CSI-Bench provides a comprehensive testbed to support this work and offers a scalable, practical resource for advancing WiFi sensing systems in healthcare and beyond. We release the full dataset and benchmark code to facilitate reproducibility and further innovation in this space.

\section*{Broader Impact}
\label{sec:broader impact}
CSI-Bench establishes a foundational step toward scalable, privacy-preserving, and contactless healthcare enabled by commodity WiFi infrastructure. By leveraging signals already emitted in everyday environments, it supports continuous health and behavioral monitoring without requiring wearables or cameras, thereby reducing barriers to adoption and ensuring dignity, inclusivity, and accessibility in care.

The benchmark supports multiple healthcare-relevant sensing capabilities: fall detection and breathing monitoring for safety and wellness tracking; activity recognition and localization for rehabilitation and behavioral analysis; user identification and proximity estimation for personalized assistance and social-interaction assessment; and motion source recognition to reduce false alarms in automated home-care systems. Collectively, these tasks lay the groundwork for comprehensive, unobtrusive smart health monitoring in homes, hospitals, and assisted living facilities.

Beyond practical healthcare applications, CSI-Bench contributes to the broader machine learning and sensing communities by providing a large-scale, in-the-wild dataset that captures the true complexity of human environments. Its diversity in users, hardware, and contexts fosters reproducible research and robust model development for real-world deployment. The high-dimensional CSI signals pose unique challenges in time-series modeling, representation learning, and domain generalization, motivating advances in trustworthy and adaptive AI systems that extend beyond healthcare.

All data are collected with consent, anonymized, and ethically managed. CSI-Bench serves both as a foundation for practical healthcare solutions and as a benchmark for high-dimensional, human-centered AI systems.

\bibliographystyle{plain}
\bibliography{refs1}


\newpage
\section*{NeurIPS Paper Checklist}

\begin{enumerate}

\item {\bf Claims}
    \item[] Question: Do the main claims made in the abstract and introduction accurately reflect the paper's contributions and scope?
    \item[] Answer: \answerYes{}
    \item[] Justification: The main claims made in the abstract and introduction accurately reflect the paper's contributions and scope.

\item {\bf Limitations}
    \item[] Question: Does the paper discuss the limitations of the work performed by the authors?
    \item[] Answer: \answerYes{} 
    \item[] Justification: Please see Section \ref{sec:limitation}.

\item {\bf Theory Assumptions and Proofs}
    \item[] Question: For each theoretical result, does the paper provide the full set of assumptions and a complete (and correct) proof?
    \item[] Answer: \answerNA{} 
    \item[] Justification: The paper does not include theoretical results. 

    \item {\bf Experimental Result Reproducibility}
    \item[] Question: Does the paper fully discLoSe all the information needed to reproduce the main experimental results of the paper to the extent that it affects the main claims and/or conclusions of the paper (regardless of whether the code and data is provided or not)?
    \item[] Answer: \answerYes{}
    \item[] Justification: The experiments are reproducible. The code, dataset and detailed instructions are provided.

\item {\bf Open access to data and code}
    \item[] Question: Does the paper provide open access to the data and code, with sufficient instructions to faithfully reproduce the main experimental results, as described in supplemental material?
    \item[] Answer: \answerYes{} 
    \item[] Justification: Please find the links to our code and dataset attached in the abstract. We have detailed instructions on how to use our data and benchmark code included. 
    \item[] Guidelines:

\item {\bf Experimental Setting/Details}
    \item[] Question: Does the paper specify all the training and test details (e.g., data splits, hyperparameters, how they were chosen, type of optimizer, etc.) necessary to understand the results?
    \item[] Answer: \answerYes{}
    \item[] Justification: The training details are discussed in Section ~\ref{app:trainingdetails}.

\item {\bf Experiment Statistical Significance}
    \item[] Question: Does the paper report error bars suitably and correctly defined or other appropriate information about the statistical significance of the experiments?
    \item[] Answer: \answerYes{}
    \item[] Justification: Results in the paper are reported with error bars representing the standard deviation across three random seeds, please see Table \ref{tab:supervised-results} and Table \ref{tab:multitask_results} in Section \ref{sec:baselines}. The primary sources of variability include random initialization, data shuffling, and train/validation/test splits. These are consistently applied to all baseline and proposed models.

\item {\bf Experiments Compute Resources}
    \item[] Question: For each experiment, does the paper provide sufficient information on the computer resources (type of compute workers, memory, time of execution) needed to reproduce the experiments?
    \item[] Answer: \answerYes{}
    \item[] Justification: The computing requirements are discussed in Appendix ~\ref{app:trainingdetails}.
    
\item {\bf Code Of Ethics}
    \item[] Question: Does the research conducted in the paper conform, in every respect, with the NeurIPS Code of Ethics \url{https://neurips.cc/public/EthicsGuidelines}?
    \item[] Answer:\answerYes{}
    \item[] Justification: Data were passively collected via commercial WiFi devices in participants’ homes, with informed consent and no direct researcher interaction. An internal ethics review addressed privacy and risk, and all data were anonymized. Participants received fair compensation.

\item {\bf Broader Impacts}
    \item[] Question: Does the paper discuss both potential positive societal impacts and negative societal impacts of the work performed?
    \item[] Answer:  \answerYes{} Please see Section \ref{sec:broader impact}.  

\item {\bf Safeguards}
    \item[] Question: Does the paper describe safeguards that have been put in place for responsible release of data or models that have a high risk for misuse (e.g., pretrained language models, image generators, or scraped datasets)?
    \item[] Answer: \answerNA{}
    \item[] Justification: The paper does not involve the release of models or data that carry a high risk for misuse or dual-use concerns. We do not release any generative models or scraped data from public sources, and the released dataset does not pose foreseeable risks that require additional safeguards.

\item {\bf Licenses for existing assets}
    \item[] Question: Are the creators or original owners of assets (e.g., code, data, models), used in the paper, properly credited and are the license and terms of use explicitly mentioned and properly respected?
    \item[] Answer: \answerYes{}
    \item[] Justification: We have cited the original papers and repositories that produced the code packages and models used in our work. For each asset, we have included the appropriate references in the paper, along with the license information and relevant URLs where applicable. All assets used are open-source and have been used in compliance with their respective licenses

\item {\bf New Assets}
    \item[] Question: Are new assets introduced in the paper well documented and is the documentation provided alongside the assets?
    \item[] Answer: \answerYes{}
    \item[] Justification: We introduce a new large-scale dataset and benchmark code suite for WiFi-based sensing tasks. Please find the links under abstract.

\item {\bf Crowdsourcing and Research with Human Subjects}
    \item[] Question: For crowdsourcing experiments and research with human subjects, does the paper include the full text of instructions given to participants and screenshots, if applicable, as well as details about compensation (if any)? 
    \item[] Answer: \answerYes{}
    \item[] Justification: Our dataset includes data from human subjects in their homes, see \ref{sec:datacollectionprotocol}, with written informed consent and fair compensation provided. Consent procedures and compensation details are summarized in the main paper, with full instructions and sample screenshots available in the supplementary material for transparency. 

\item {\bf Institutional Review Board (IRB) Approvals or Equivalent for Research with Human Subjects}
    \item[] Question: Does the paper describe potential risks incurred by study participants, whether such risks were discLoSed to the subjects, and whether Institutional Review Board (IRB) approvals (or an equivalent approval/review based on the requirements of your country or institution) were obtained?
    \item[] Answer: \answerYes{}
    \item[] Justification: Our research involved passive data collection in participants’ homes using commercial WiFi devices, with no direct interaction or intervention, see \ref{sec:datacollectionprotocol}.

\end{enumerate}

\newpage

\appendix
\section{Dataset Description}
\label{app:dataset}

This appendix details the composition and collection protocols of CSI-Bench.

\subsection{Subjects and Scenarios}
CSI-Bench includes CSI data from 35 individual users (U01–U35), comprising 26 males and 9 females aged between 23 and 42 years, with heights ranging from 155 to 185 cm and body weights from 45 to 90 kg. In addition, six two-user sessions (UM01–UM06) are recorded to support multi-person interaction analysis. To further diversify subject types, the dataset includes 20 pets (P01–P20), with body weights from 6 to 40 kg, and a dedicated two-pet scenario (PM01). Finally, four distinct fan-based motion scenes (F01–F04) capture ambient signal patterns caused by oscillating and ceiling fans.

\subsection{Environments}
\label{app:environments}
Data is collected across 26 distinct real-world environments (E01–E26), including studio apartments, multi-bedroom apartments, townhouses, and multi-floor single-family houses. These environments vary in layout complexity, room geometry, wall materials, and furniture density, introducing rich multipath and occlusion effects. A summary of all environments is provided in Table~\ref{tab:env-summary}.

\subsection{Devices and Hardware Diversity}
\begin{figure*}[h]
    \centering
    \includegraphics[width=0.68\linewidth]{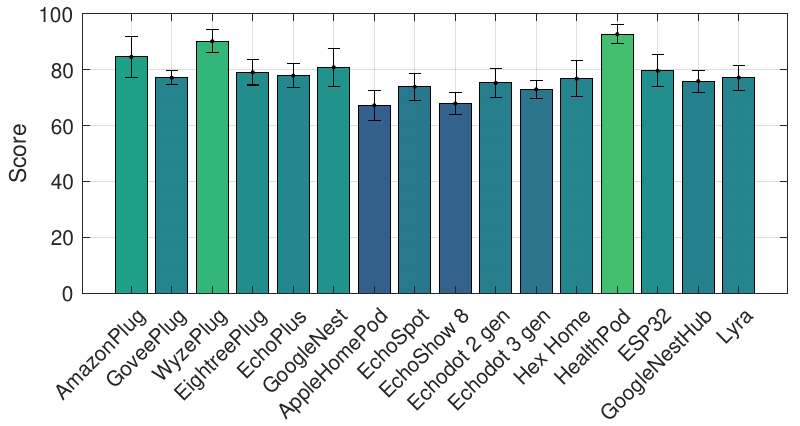}
    \caption{Average CSI quality scores of 16 widely used IoT devices evaluated using our CSI verification tool. Each bar represents the mean score across five measurement trials, with error bars indicating the standard deviation.}
    \label{fig:iot_score}
\end{figure*}

To ensure broad coverage of real-world IoT infrastructure, we select 16 types of commercial WiFi-enabled edge devices operating across both 2.4 GHz and 5 GHz bands, with bandwidths of 20, 40, and 80 MHz. Devices span major chipset vendors such as Qualcomm, Broadcom, Espressif, and NXP, covering configurations from low-cost smart plugs to high-performance routers and smart speakers. Prior to data collection, each device is evaluated using our in-house CSI verification tool to assess signal consistency, sampling stability, and amplitude dynamics. Devices that pass quality thresholds are used in deployment. Figure~\ref{fig:iot_score} presents the CSI quality scores across candidate devices; Table~\ref{tab:device_summary} lists their specifications, together with average RSSI values (in dBm) measured over one-hour static indoor sessions for each device. These RSSI readings characterize real-world signal strength and serve as a practical proxy for transmit-power variability across hardware families.

\subsection{Task-Specific Dataset Statistics}
CSI-Bench supports both single-task specialist datasets and a co-labeled multi-task dataset. Task-wise breakdowns include the number of samples, users, environments, and devices, as detailed in Table~\ref{tab:task-summary-extended}. Each task is annotated with appropriate labels to support supervised learning, multi-task training, and cross-domain evaluation.

\paragraph{Note on Evaluation Splits.}
For rigorous benchmarking, CSI-Bench defines task-specific evaluation splits based on difficulty levels (Easy, Medium, Hard) and out-of-distribution (OOD) axes (cross-user, cross-environment, cross-device). These splits are introduced in Sections~\ref{app:fall}–\ref{app:multi-task} for each task and are used to generate the experimental results reported in Appendix~\ref{app:results}.

\setlength{\tabcolsep}{3pt}
\renewcommand{\arraystretch}{2}

\begin{table}[t]
\centering
\small
\caption{Summary of tasks and dataset statistics.}
\label{tab:task-summary-extended}
\begin{tabular}{>{\raggedright\arraybackslash}p{1.2cm} >{\raggedright\arraybackslash}p{3 cm} >{\raggedright\arraybackslash}p{2.2cm} >{\raggedright\arraybackslash}p{1.5cm} >{\raggedright\arraybackslash}p{1.5cm} >{\raggedright\arraybackslash}p{1.7cm} >{\raggedright\arraybackslash}p{1.65cm}}
\toprule
\textbf{Task} & \textbf{Users} & \textbf{Envs} & \textbf{Gender} & \textbf{Age (yrs)} & \textbf{Height (cm)} & \textbf{Weight (kg)} \\
\midrule
Fall & U06 - U22 & E21 - E26 & 14M / 3F & 23 - 42 & 156 - 182 & 46 - 90 \\
Breath & U06, U23, U24 & E05, E09, E10 & 1M / 2F & 27 - 32 & 160 - 173 & 60 - 88 \\
Loc. & \makecell[l]{U01, U05, U06\\ UM01 - UM05} & \makecell[l]{E01, E03, E04\\ E06 - E08} & 4M / 4F & 27 - 41 & 155 - 175 & 48 - 90 \\
Prox. & U01 - U06 & E01 - E06 & 2M / 4F & 26 - 41 & 163 - 173 & 45 - 90 \\
HAR & U01 - U06 & E01 - E06 & 2M / 4F & 26 - 41 & 163 - 173 & 45 - 90 \\
UID & U01 - U06 & E01 - E06 & 2M / 4F & 26 - 41 & 163 - 173 & 45 - 90 \\

\multirow{3}{*}{MSR} 
& P01 - P20, PM01 & \makecell[l]{E11 - E13,\\ E15 - E20} & - & - & - & 6 - 40\\
& \makecell[l]{U03 - U04, U08 - U10,\\ U13, U18, U20,\\ U25 - U27, U29 - U35,\\ UM06} & E11 - E20 & 15M / 3F & 23 - 35 & 155 - 185 & 50 - 90 \\
& F01 - F04  & E11 - E13 & - & - & - & -\\
\bottomrule
\end{tabular}
\end{table}

\renewcommand{\arraystretch}{1.3}
\setlength{\tabcolsep}{3pt}

\begin{table}[h]
\centering
\small
\caption{Summary of edge devices, WiFi chipsets, and specifications.}
\label{tab:device_summary}
\begin{tabular}{
>{\raggedright\arraybackslash}p{2.2cm}
>{\raggedright\arraybackslash}p{1.6cm}
>{\raggedright\arraybackslash}p{2.2cm}
>{\centering\arraybackslash}p{1.2cm}
>{\centering\arraybackslash}p{1.8cm}
>{\centering\arraybackslash}p{1.7cm}
>{\centering\arraybackslash}p{1.7cm}}
\toprule
\textbf{Device} & \textbf{Chipset} & \textbf{Model} & \textbf{Antenna} & \textbf{Bandwidth} & \textbf{Band} & \textbf{RSSI} \\
\midrule
AmazonPlug & MediaTek & MT7697N & 1x1 & 20MHz & 2.4G & -49.95 dBm \\
GoveePlug & Espressif & ESP8266/ESP8285 & 1x1 & 20MHz & 2.4G & -50.26 dBm \\
WyzePlug & Espressif & ESP8266/ESP8285 & 1x1 & 20MHz & 2.4G & -55.57 dBm \\
EightreePlug & Espressif & ESP8266/ESP8285 & 1x1 & 20MHz & 2.4G & -54.01 dBm \\
EchoPlus & MediaTek & MT8516 & 1x1 & 20/40/80MHz & 2.4G \& 5G & -42.24 dBm \\
GoogleNest & Qualcomm & IPQ4019 & 1x1 & 20/40MHz & 2.4G \& 5G & -49.41 dBm \\
AppleHomePod & -- & -- & 1x1 & 20/40MHz & 2.4G \& 5G & -52.56 dBm \\
EchoSpot & MediaTek & MT6625L & 1x1 & 20/40MHz & 2.4G \& 5G & -51.20 dBm \\
EchoShow 8 & MediaTek & MT8183 & 1x1 & 20/40/80MHz & 2.4G \& 5G & -38.14 dBm \\
Echodot 2 gen & MediaTek & MT6625LN & 1x1 & 20/40MHz & 2.4G \& 5G & -76.83 dBm \\
Echodot 3 gen & MediaTek & MT7658CSN & 1x1 & 20/40/80MHz & 2.4G \& 5G & -57.75 dBm \\
Hex Home & Qualcomm & -- & 1x2 & 20/40MHz & 5G & -- \\
HealthPod & NXP & 88W889 & 2x2 & 20/40/80MHz & 5G & -- \\
ESP32 & Espressif & S3 & 1x1 & 20/40MHz & 2.4G & -62.86 dBm \\
GoogleNestHub & Broadcom & BCM4345 & 1x1 & 20/40/80MHz & 2.4G \& 5G & -60.39 dBm \\
Lyra & Qualcomm & -- & 2x2 & 20/40/80MHz & 2.4G \& 5G & -- \\
\bottomrule
\end{tabular}
\end{table}

\begin{table}[t]
\centering
\small
\caption{Summary of environments (\textit{XBXB} indicates \textit{X bedrooms and X bathrooms}).}
\label{tab:env-summary}
\begin{tabular}{>{\raggedright\arraybackslash}p{1.2cm} >{\raggedright\arraybackslash}p{3.2cm} >{\centering\arraybackslash}p{2cm} >{\raggedright\arraybackslash}p{2.5cm} >{\centering\arraybackslash}p{1.5cm}}
\toprule
\textbf{Env ID} & \textbf{Type} & \textbf{Area (sqft)} & \textbf{Layout Type} & \textbf{\# Floors} \\
\midrule
E01 & Single-family house   & 2400 & Multi-room & 3 \\
E02 & Apartment              & 633 & Studio & 1 \\
E03 & Apartment              & 1077  & 2B2B & 1 \\
E04 & Apartment              & 790 & 1B1B & 1 \\
E05 & Apartment              & 714 & 1B1B & 1 \\
E06 & Apartment              & 1652 & 3B2B & 1 \\
E07 & Apartment              & 1250  & 2B2B & 1 \\
E08 & Single-family house    & 1790 & Multi-room & 2 \\
E09 & Apartment              & 1200  & 2B2B & 1 \\
E10 & Single-family house    & 1904 & Multi-room & 2 \\
E11 & Single-family house    & 1352 & Multi-room & 2 \\
E12 & Apartment              & 830 & 1B1B & 1 \\
E13 & Apartment              & 2242  & 4B2B & 1 \\
E14 & Single-family house    & 1700 & Multi-room & 2 \\
E15 & Single-family house    & 2000 & Multi-room & 2 \\
E16 & Apartment              & 960  & 1B1B & 1 \\
E17 & Apartment              & 860  & 1B1B & 1 \\
E18 & Single-family house    & 1680 & Multi-room & 2 \\
E19 & Town house             & 2600 & Multi-room & 4 \\
E20 & Office                 & 1224  & Partitioned rooms & 1 \\
E21 & Apartment              & 700 & 2B1B & 1 \\
E22 & Single-family house    & 1300  & Multi-room & 2 \\
E23 & Office                 & 1500 &  Partitioned rooms & 1 \\
E24 & Single-family house    & 1250 & Multi-room & 2 \\
E25 & Single-family house    & 1400 & Multi-room & 2 \\
E26 & Single-family house    & 900  & Multi-room & 3 \\
\bottomrule
\end{tabular}
\end{table}

\subsection{Fall Detection}
\label{app:fall}
The Fall Detection dataset is designed to evaluate human fall recognition in real residential settings using commodity WiFi hardware. Data is collected with synchronized video ground-truth under varied hardware and environmental conditions.

\textbf{Subjects and Scenarios.}
The dataset includes 17 participants across 6 indoor environments. Activities include casual walking, sitting, lying down, and falling. Scenarios include both LoS and NLoS layouts, with added noise from ambient sources such as ceiling fans to simulate realistic deployments.

\textbf{Hardware Setup.}
WiFi CSI data is primarily collected using NXP88W8997 2×2 802.11ac chipsets operating at 5.18 GHz with a 40 MHz bandwidth. Each transmitter-receiver pair forms 4 spatial links and records 58 subcarriers at a sampling rate of 100 Hz. Additionally, a smaller portion of the data is collected using ESP32-S3 devices, which operate at 2.4 GHz with a 1×1 antenna setup and capture 64 subcarriers.

\textbf{Data Collection Protocol.}
Each session lasts 1–5 minutes, capturing both routine and fall-related activities. Fall events are annotated using synchronized video recordings.

\textbf{Scale and Composition:}
6 environments (homes and offices); 17 participants; 2,770 fall events; 3,930 non-fall activities.

\textbf{Difficulty-Level Evaluation.}
Test samples are stratified into \textit{Easy}, \textit{Medium}, and \textit{Hard} tiers based on environmental complexity and device quality. Medium includes fan-induced interference; Hard includes ESP32-based low-quality CSI.

\subsection{Breathing Detection}

This dataset captures subtle respiration signals under natural sleep conditions using diverse IoT hardware in real homes.

\textbf{Subjects and Scenarios.}
Breathing data is collected from 3 participants across 3 residential environments. Deployment setups range from same-room (LoS) to cross-room (NLoS), with and without fan interference.

\textbf{Hardware Setup.}
Devices include Amazon Echo Dots, Echo Plus, Google Nest Hub, and Qualcomm-based 5 GHz routers. Sampling is fixed at 30 Hz.

\textbf{Data Collection Protocol.}
Overnight sessions are passively recorded during natural sleep without intervention. Participants optionally log activity context.

\textbf{Scale and Composition:} $\sim$55,000 breathing samples; $\sim$45,000 empty-room samples; $\sim$11,400 fan-interfered samples; Diverse device placements and heights (0.47–2.18 m).

\textbf{Difficulty-Level Evaluation.}
Difficulty is assigned based on device-user distance, interference level, and deployment complexity. Hard tiers involve distant NLoS setups and overlapping fan motion.

\subsection{Room-Level Localization}

This dataset supports room-level user localization in typical households with both single- and multi-user presence.

\textbf{Subjects and Scenarios.}
Data is collected from 8 users in 6 homes. Three rooms per home are labeled for occupancy. Scenarios include both single and two-user activity.

\textbf{Hardware Setup.}
Devices span 8 types (Echo, Google Nest, Apple HomePod, etc.) operating on 2.4/5 GHz at 30 or 100 Hz. Bandwidths vary from 20–80 MHz.

\textbf{Data Collection Protocol.}
Users annotate their room presence and co-occupancy manually. Sessions reflect natural daily activities.

\textbf{Scale and Composition:}
3,805 single-user samples;
3,257 multi-user samples;
6 diverse environments;
8 device types.

\textbf{Difficulty-Level Evaluation.}
Tiers are defined by user count and hardware quality. Easy cases use high-quality CSI from 5 GHz devices; hard cases include 2.4 GHz plugs and multi-user ambiguity.

\subsection{Motion Source Recognition}

This dataset captures motion patterns from humans, pets, robots, and fans in diverse indoor settings.

\textbf{Subjects and Scenarios.}
Data include 13 humans (ages 23–34), 11 pets, Roomba robots, and oscillating fans. Activities include walking, sneaking, and simulated intrusion. Environments span homes, townhouses, and offices.

\textbf{Hardware Setup.}
CSI is collected via NXP88W8997 2×2 devices at 100 Hz over 58 subcarriers.

\textbf{Data Collection Protocol.}
Each session lasts 3–8 minutes. Human data is optionally logged by users; non-human motion is passively captured.

\textbf{Scale and Composition:}
$\sim$150K seconds of human motion;
$\sim$2,000 minutes of pet activity;
$\sim$1,000 minutes of robot activity;
$\sim$200 minutes of fan motion.

\textbf{Difficulty-Level Evaluation.}
Difficulty is based on motion type, subject diversity, and signal quality. Easy cases include clean human walking or small pets; hard cases include multi-subjects, large pets, or intrusion patterns under NLoS.

\subsection{Multi-task Dataset}
\label{app:multi-task}
This dataset enables multi-task learning across activity recognition, user identification, and proximity estimation.

\textbf{Subjects and Scenarios.}
Six users perform 5 activities across 6 homes: walking (at 4 distances), running, jumping, seated breathing, and waving. Cross-user and cross-environment samples are included.

\textbf{Hardware Setup.}
Each environment uses 5–7 IoT devices across 2.4/5 GHz bands. Devices include Echo, Google Nest, Apple HomePod, ESP32 plugs, and more.

\textbf{Data Collection Protocol.}
Each activity lasts 3–6 minutes. Participants use a lightweight UI to annotate activity boundaries and proximity distances.

\textbf{Scale and Composition:}
41,503 total samples;
$\sim$5,000–6,000 samples per activity;
4 proximity distances: 0.5, 1.5, 2.5, 3.5 m.

\textbf{Cross-Domain Evaluation.}
To evaluate generalization, held-out domains include:
\begin{itemize}
\item \textbf{Cross-User:} U02
\item \textbf{Cross-Environment:} E05
\item \textbf{Cross-Device:} Amazon Plug, Echo Spot
\end{itemize}

These exclusions are reserved for OOD test sets used in Appendix~\ref{app:results}.

\subsection{Annotation Tool}
\label{app:annotation_tool}
To facilitate user-friendly and accurate labeling during in-the-wild data collection, we developed a lightweight annotation tool based on Google Spreadsheets for accessibility and cross-platform compatibility.

\begin{figure*}[h]
    \centering
    \includegraphics[width=0.95\linewidth]{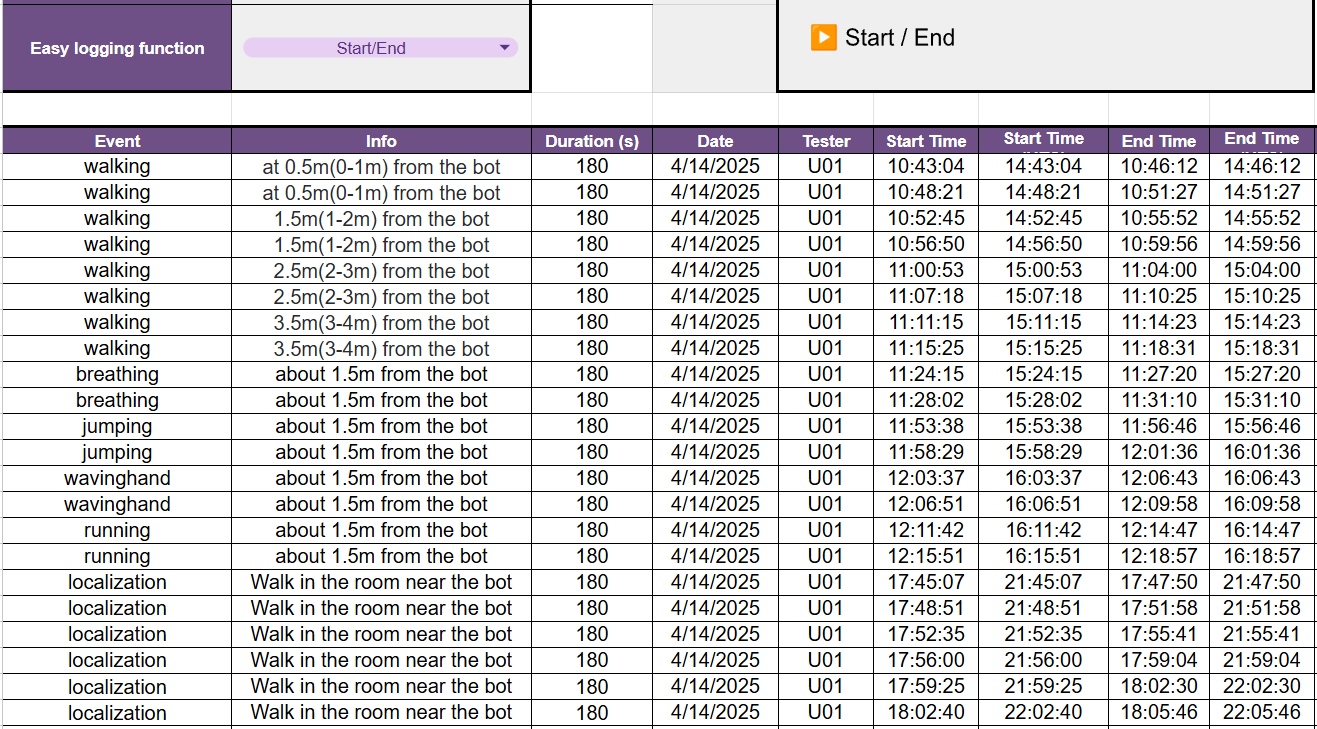}
    \caption{Screenshot of the Google Spreadsheet–based annotation tool used by participants to record activities and timestamps during data collection.}
    \label{fig:annotation_tool}
\end{figure*}

As illustrated in Figure \ref{fig:annotation_tool}, the tool provides a simple interface where users can log activities through ``Start/End'' button clicks corresponding to predefined motion types (e.g., walking, breathing, jumping, waving hand, running, localization). Each button click automatically records the timestamp, activity label, tester ID, and session duration, ensuring precise temporal alignment with the collected CSI data.

The design emphasizes ease of use and minimal user burden. Participants simply tap the relevant button when starting and finishing an activity—no manual typing or complex input is required. The captured logs include information such as activity type, approximate user–device distance, location context, and timestamps, which are later aligned with the CSI files through automated scripts described in Section \ref{subsec:continuous_data_recording}.

During the collection campaign, participants typically logged around three minutes per activity type per day, covering multiple motion categories. While logging was optional, this structured yet flexible protocol ensured sufficient labeled samples for model training while allowing users to behave naturally in their environments.

\section{Model Architectures and Training Details}
\label{app:model-details}
To support rigorous evaluation across in-distribution, cross-domain, and few-shot generalization, we implement and benchmark a suite of neural network models representative of contemporary time-series and vision-inspired architectures. All models are implemented in PyTorch and trained under consistent protocols unless otherwise noted.
\subsection{Supervised Learning Architectures}
We benchmark the following supervised architectures across all tasks:

\paragraph{Multi-Layer Perceptron (MLP)}
The MLP model consists of three fully-connected layers with ReLU activations and dropout for regularization. The input to the model is a flattened CSI feature vector, capped at a maximum of 10,000 dimensions to control memory usage. Specifically, the architecture is: [Input → Linear(512) → ReLU → Dropout(0.5) → Linear(128) → ReLU → Dropout(0.3) → Linear(Output classes)].

\paragraph{Long Short-Term Memory (LSTM)}
Our LSTM baseline includes a bidirectional LSTM with two layers, each containing 256 hidden units. A linear classifier follows this, accompanied by dropout for regularization: [Input → Bi-LSTM(256, 2 layers, dropout=0.3) → Linear(256) → ReLU → Dropout(0.3) → Linear(Output classes)].

\paragraph{ResNet-18}
We modify a standard ResNet-18 architecture to accept single-channel input (WiFi CSI data) by adapting the first convolutional layer accordingly. The final fully connected layer is tailored to the task-specific number of classes.

\paragraph{Vision Transformer (ViT)}
The ViT model converts input CSI data into embedded patches using convolutional patch embeddings, followed by Transformer encoder layers (6 layers, embedding dimension=128, and 4 heads). A class token is prepended for classification tasks. Dropout and layer normalization are employed for stability.

\paragraph{Transformer}
This architecture employs Transformer encoder layers (4 layers, model dimension=256, 8 attention heads). Inputs are linearly projected into the model dimension, positional encodings are added, and global average pooling is applied before classification. Dropout is set to 0.1 to prevent overfitting.

\paragraph{PatchTST}
PatchTST utilizes temporal patch embeddings (patch length=16, stride=8) processed through Transformer encoder layers (4 layers, embedding dimension=128, 4 heads). The architecture includes positional encodings, dropout (0.1), and a CLS token or mean pooling strategy for final prediction.

\paragraph{TimeSformer-1D}
TimeSformer-1D adopts patch embeddings (patch size=4) followed by separate temporal and feature attention within Transformer blocks (4 layers, embedding dimension=128, 8 heads). A class token and positional embeddings are included for classification, with dropout layers added for robustness.

All models use a final linear classifier and are initialized using Xavier uniform initialization unless otherwise specified.

\subsection{Multi-Task Learning with Adapters}
To enable efficient multi-task learning across diverse WiFi sensing tasks, we implement task-specific adapter modules on top of a shared backbone:
\begin{itemize}
    \item \textbf{LoRA Adapters:} For Transformer backnone model, we apply LoRA to the attention modules. Each task has separate adapter weights (rank=8, $\alpha$=32, dropout=0.05).
    \item \textbf{Task Adapters:} A residual two-layer bottleneck MLP (down-project, GELU, up-project, followed by LayerNorm) is applied post-backbone for each task.
    \item \textbf{Task-Specific Heads:} Each task has a separate classification head, initialized via Xavier uniform.
\end{itemize}

During training, we activate one task at a time and update both the shared backbone and the active task’s adapter and head.

\subsection{Training Protocol}
\label{app:trainingdetails}
All models are trained with the AdamW optimizer, a batch size of 128, and initial learning rate of $1e-3$. We apply cosine learning rate decay with 5 warm-up epochs and weight decay of $1e-5$. Training lasts up to 100 epochs, with early stopping based on validation loss (patience = 15). We use categorical cross-entropy as the loss function. The hyperparameter are tuned based on models' accuracy on validation dataset. Data is loaded from HDF5 using standardized splits as discussed in Section~\ref{sec:eval-protocols} and label mappings. Our experiments utilize NVIDIA GeForce RTX 4090 GPUs and AWS Sagemaker involved training with three random seeds across all datasets. For training in AWS Sagemaker, we use ml.g5.g5.12xlarge, which includes 4 NVIDIA A10G Tensor Core GPUs. The training time for tasks ranges from 0.5 hour to 13 hours.

\section{Additional Experiments}
\label{app:results}
The results in Appendix~\ref{app:results} are stratified by the difficulty tiers and OOD evaluation protocols defined in Appendix A.1–A.5. For each task, performance is reported across (i) three difficulty levels (Easy, Medium, Hard) reflecting environmental and signal complexity for single-task datasets (Appendix~\ref{app:difficulty-results}), and (ii) three out-of-distribution (OOD) axes—cross-user, cross-environment, and cross-device—for multi-task datasets (Appendix~\ref{app:difficulty-results}). All splits are predefined during data collection and are described per task in Appendix~\ref{app:dataset}.

\subsection{Evaluation with Difficulty Tiers}
\label{app:difficulty-results}

Table~\ref{tab:falldetection} compares \textbf{Fall Detection} performance across three difficulty levels. All models perform well under the Easy setting, with LSTM, PatchTST, ResNet18, and ViT achieving F1-scores above 97\%. MLP underperforms due to limited temporal modeling. In the Medium tier, performance drops notably—ResNet18 and ViT remain strong (F1 $\sim$77\%), while PatchTST degrades significantly (F1 $\sim$56\%). TimeSformer-1D and Transformer show moderate results. In the Hard tier, ResNet18 leads with 68.08\% F1, while others degrade further. The larger variance in Medium and Hard tiers is due to smaller dataset sizes, which increase sensitivity to noise and reduce performance stability.

\begin{table}[t]
\centering
\caption{ \textbf{Fall Detection} performance comparison of supervised models. Accuracy (Acc) and F1-score are reported as mean ± std (\%) over three runs.}
\label{tab:falldetection}
\small
\begin{tabular}{l cc cc cc}
\toprule
\multirow{2}{*}{\textbf{Model}} & \multicolumn{2}{c}{\textbf{Easy}} & \multicolumn{2}{c}{\textbf{Medium}} & \multicolumn{2}{c}{\textbf{Hard}} \\
\cmidrule(lr){2-3} \cmidrule(lr){4-5} \cmidrule(lr){6-7}
& Acc & F1 & Acc & F1 & Acc & F1 \\
\midrule
LSTM~\cite{LSTM1997} & 97.62 \textpm {\scriptsize 0.52} & 97.62 \textpm {\scriptsize 0.52} & 69.12 \textpm {\scriptsize 5.63} & 68.20 \textpm {\scriptsize 5.19} & 67.12 \textpm {\scriptsize 2.96} & 66.05 \textpm {\scriptsize 4.03} \\
MLP~\cite{MLP} & 94.84 \textpm {\scriptsize 0.85} & 94.84 \textpm {\scriptsize 0.85} & 70.59 \textpm {\scriptsize 9.61} & 70.19 \textpm {\scriptsize 9.91} & 63.70 \textpm {\scriptsize 2.37} & 63.41 \textpm {\scriptsize 2.25} \\
PatchTST~\cite{PatchTST2023} & 97.13 \textpm {\scriptsize 0.72} & 97.13 \textpm {\scriptsize 0.72} & 61.76 \textpm {\scriptsize 7.59} & 56.31 \textpm {\scriptsize 12.22} & 62.67 \textpm {\scriptsize 2.82} & 61.36 \textpm {\scriptsize 4.14} \\
ResNet18~\cite{kaiming2016} & 97.27 \textpm {\scriptsize 0.32} & 97.27 \textpm {\scriptsize 0.32} & 77.94 \textpm {\scriptsize 5.63} & 76.96 \textpm {\scriptsize 6.46} & 68.84 \textpm {\scriptsize 3.04} & 68.08 \textpm {\scriptsize 3.58} \\
TimeSformer-1D~\cite{TimeSformer2021} & 96.58 \textpm {\scriptsize 0.50} & 96.59 \textpm {\scriptsize 0.49} & 67.65 \textpm {\scriptsize 7.59} & 64.55 \textpm {\scriptsize 11.72} & 65.75 \textpm {\scriptsize 9.29} & 61.19 \textpm {\scriptsize 17.16} \\
Transformer~\cite{attention2017} & 97.08 \textpm {\scriptsize 0.54} & 97.08 \textpm {\scriptsize 0.54} & 69.12 \textpm {\scriptsize 5.63} & 68.10 \textpm {\scriptsize 5.91} & 65.07 \textpm {\scriptsize 6.94} & 63.89 \textpm {\scriptsize 7.40} \\
ViT~\cite{ViT2021} & 97.40 \textpm {\scriptsize 0.42} & 97.40 \textpm {\scriptsize 0.42} & 77.94 \textpm {\scriptsize 13.04} & 77.07 \textpm {\scriptsize 14.46} & 65.75 \textpm {\scriptsize 3.71} & 64.06 \textpm {\scriptsize 6.66} \\
\bottomrule
\end{tabular}
\vspace{1ex}
\end{table}

\begin{table}[H]
\centering
\caption{ \textbf{Breathing Detection} performance comparison of supervised models. Accuracy (Acc) and F1-score are reported as mean ± std (\%) over three runs.}
\label{tab:breathingdetection_subset}
\small
\begin{tabular}{l cc cc cc}
\toprule
\multirow{2}{*}{\textbf{Model}} & \multicolumn{2}{c}{\textbf{Easy}} & \multicolumn{2}{c}{\textbf{Medium}} & \multicolumn{2}{c}{\textbf{Hard}} \\
\cmidrule(lr){2-3} \cmidrule(lr){4-5} \cmidrule(lr){6-7}
& Acc & F1 & Acc & F1 & Acc & F1 \\
\midrule
LSTM~\cite{LSTM1997} & 99.11 \textpm {\scriptsize 0.17} & 99.11 \textpm {\scriptsize 0.17} & 98.61 \textpm {\scriptsize 0.13} & 98.61 \textpm {\scriptsize 0.13} & 98.08 \textpm {\scriptsize 0.28} & 98.08 \textpm {\scriptsize 0.28} \\
MLP~\cite{MLP} & 98.54 \textpm {\scriptsize 0.14} & 98.54 \textpm {\scriptsize 0.14} & 97.67 \textpm {\scriptsize 0.15} & 97.67 \textpm {\scriptsize 0.15} & 96.46 \textpm {\scriptsize 0.13} & 96.46 \textpm {\scriptsize 0.13} \\
PatchTST~\cite{PatchTST2023} & 99.20 \textpm {\scriptsize 0.06} & 99.20 \textpm {\scriptsize 0.06} & 98.77 \textpm {\scriptsize 0.19} & 98.77 \textpm {\scriptsize 0.19} & 98.49 \textpm {\scriptsize 0.22} & 98.49 \textpm {\scriptsize 0.22} \\
ResNet18~\cite{kaiming2016} & 98.94 \textpm {\scriptsize 0.17} & 98.94 \textpm {\scriptsize 0.17} & 98.42 \textpm {\scriptsize 0.16} & 98.42 \textpm {\scriptsize 0.16} & 98.32 \textpm {\scriptsize 0.25} & 98.32 \textpm {\scriptsize 0.25} \\
TimeSformer-1D~\cite{TimeSformer2021} & 99.05 \textpm {\scriptsize 0.22} & 99.05 \textpm {\scriptsize 0.22} & 98.29 \textpm {\scriptsize 0.31} & 98.29 \textpm {\scriptsize 0.31} & 98.60 \textpm {\scriptsize 0.23} & 98.60 \textpm {\scriptsize 0.23} \\
Transformer~\cite{attention2017} & 98.23 \textpm {\scriptsize 0.24} & 98.23 \textpm {\scriptsize 0.24} & 97.31 \textpm {\scriptsize 0.47} & 97.31 \textpm {\scriptsize 0.47} & 97.54 \textpm {\scriptsize 0.31} & 97.54 \textpm {\scriptsize 0.31} \\
ViT~\cite{ViT2021} & 99.56 \textpm {\scriptsize 0.08} & 99.56 \textpm {\scriptsize 0.08} & 99.41 \textpm {\scriptsize 0.08} & 99.41 \textpm {\scriptsize 0.08} & 99.17 \textpm {\scriptsize 0.11} & 99.17 \textpm {\scriptsize 0.11} \\
\bottomrule
\end{tabular}
\vspace{1ex}
\end{table}

Table~\ref{tab:breathingdetection_subset} presents breathing detection results, where all models maintain high accuracy and F1-scores ($>$96\%) across tiers. ViT performs best, achieving over 99\% F1 consistently. LSTM and PatchTST follow closely, especially in the Easy setting. Even in the Hard tier, model performance drops only slightly. ResNet18 and TimeSformer-1D also generalize well, with minimal performance variance. The results suggest that breathing patterns are relatively easier to model and robust to environmental changes.

\begin{table}[t]
\centering
\caption{\textbf{Localization} performance comparison of supervised models. Accuracy (Acc) and F1-score are reported as mean ± std (\%) over three runs.}
\label{tab:localization}
\small
\begin{tabular}{l cc cc cc}
\toprule
\multirow{2}{*}{\textbf{Model}} & \multicolumn{2}{c}{\textbf{Easy}} & \multicolumn{2}{c}{\textbf{Medium}} & \multicolumn{2}{c}{\textbf{Hard}} \\
\cmidrule(lr){2-3} \cmidrule(lr){4-5} \cmidrule(lr){6-7}
& Acc & F1 & Acc & F1 & Acc & F1 \\
\midrule
LSTM~\cite{LSTM1997} & 99.72 \textpm {\scriptsize 0.32} & 99.75 \textpm {\scriptsize 0.29} & 100.00 \textpm {\scriptsize 0.00} & 100.00 \textpm {\scriptsize 0.00} & 98.31 \textpm {\scriptsize 0.50} & 98.31 \textpm {\scriptsize 0.50} \\
MLP~\cite{MLP} & 91.36 \textpm {\scriptsize 0.93} & 92.03 \textpm {\scriptsize 0.82} & 96.11 \textpm {\scriptsize 1.31} & 96.18 \textpm {\scriptsize 1.29} & 80.20 \textpm {\scriptsize 1.06} & 80.03 \textpm {\scriptsize 1.19} \\
PatchTST~\cite{PatchTST2023} & 100.00 \textpm {\scriptsize 0.00} & 100.00 \textpm {\scriptsize 0.00} & 99.90 \textpm {\scriptsize 0.19} & 99.95 \textpm {\scriptsize 0.10} & 99.86 \textpm {\scriptsize 0.17} & 99.86 \textpm {\scriptsize 0.18} \\
ResNet18~\cite{kaiming2016} & 100.00 \textpm {\scriptsize 0.00} & 100.00 \textpm {\scriptsize 0.00} & 100.00 \textpm {\scriptsize 0.00} & 100.00 \textpm {\scriptsize 0.00} & 100.00 \textpm {\scriptsize 0.00} & 100.00 \textpm {\scriptsize 0.00} \\
TimeSformer-1D~\cite{TimeSformer2021} & 100.00 \textpm {\scriptsize 0.00} & 100.00 \textpm {\scriptsize 0.00} & 100.00 \textpm {\scriptsize 0.00} & 100.00 \textpm {\scriptsize 0.00} & 100.00 \textpm {\scriptsize 0.00} & 100.00 \textpm {\scriptsize 0.00} \\
Transformer~\cite{attention2017} & 99.30 \textpm {\scriptsize 0.36} & 99.40 \textpm {\scriptsize 0.24} & 99.90 \textpm {\scriptsize 0.19} & 99.90 \textpm {\scriptsize 0.19} & 98.95 \textpm {\scriptsize 0.66} & 98.95 \textpm {\scriptsize 0.66} \\
ViT~\cite{ViT2021} & 99.79 \textpm {\scriptsize 0.42} & 99.82 \textpm {\scriptsize 0.35} & 99.90 \textpm {\scriptsize 0.19} & 99.90 \textpm {\scriptsize 0.19} & 99.50 \textpm {\scriptsize 0.23} & 99.50 \textpm {\scriptsize 0.23} \\
\bottomrule
\end{tabular}
\vspace{1ex}
\end{table}

\begin{table}[t]
\centering
\caption{\textbf{Motion Source Recognition} performance comparison of supervised models. Accuracy (Acc) and F1-score are reported as mean ± std (\%) over three runs.}
\label{tab:motionsourcerecognition}
\small
\begin{tabular}{l cc cc cc}
\toprule
\multirow{2}{*}{\textbf{Model}} & \multicolumn{2}{c}{\textbf{Easy}} & \multicolumn{2}{c}{\textbf{Medium}} & \multicolumn{2}{c}{\textbf{Hard}} \\
\cmidrule(lr){2-3} \cmidrule(lr){4-5} \cmidrule(lr){6-7}
& Acc & F1 & Acc & F1 & Acc & F1 \\
\midrule
LSTM~\cite{LSTM1997} & 96.65 \textpm {\scriptsize 0.96} & 96.99 \textpm {\scriptsize 0.78} & 98.79 \textpm {\scriptsize 0.11} & 98.80 \textpm {\scriptsize 0.11} & 96.94 \textpm {\scriptsize 0.94} & 96.94 \textpm {\scriptsize 0.95} \\
MLP~\cite{MLP} & 98.21 \textpm {\scriptsize 0.28} & 98.29 \textpm {\scriptsize 0.18} & 99.13 \textpm {\scriptsize 0.11} & 99.13 \textpm {\scriptsize 0.11} & 98.19 \textpm {\scriptsize 0.36} & 98.19 \textpm {\scriptsize 0.36} \\
PatchTST~\cite{PatchTST2023} & 98.01 \textpm {\scriptsize 0.69} & 98.28 \textpm {\scriptsize 0.54} & 98.59 \textpm {\scriptsize 0.36} & 98.59 \textpm {\scriptsize 0.36} & 97.49 \textpm {\scriptsize 0.71} & 97.49 \textpm {\scriptsize 0.72} \\
ResNet18~\cite{kaiming2016} & 99.86 \textpm {\scriptsize 0.11} & 99.86 \textpm {\scriptsize 0.11} & 99.73 \textpm {\scriptsize 0.05} & 99.73 \textpm {\scriptsize 0.05} & 99.48 \textpm {\scriptsize 0.32} & 99.48 \textpm {\scriptsize 0.32} \\
TimeSformer-1D~\cite{TimeSformer2021} & 96.56 \textpm {\scriptsize 0.64} & 96.92 \textpm {\scriptsize 0.56} & 98.68 \textpm {\scriptsize 0.18} & 98.69 \textpm {\scriptsize 0.18} & 97.32 \textpm {\scriptsize 0.31} & 97.31 \textpm {\scriptsize 0.32} \\
Transformer~\cite{attention2017} & 98.73 \textpm {\scriptsize 0.62} & 98.80 \textpm {\scriptsize 0.55} & 98.63 \textpm {\scriptsize 0.17} & 98.63 \textpm {\scriptsize 0.17} & 98.08 \textpm {\scriptsize 0.55} & 98.08 \textpm {\scriptsize 0.55} \\
ViT~\cite{ViT2021} & 98.38 \textpm {\scriptsize 0.87} & 98.41 \textpm {\scriptsize 0.81} & 99.27 \textpm {\scriptsize 0.32} & 99.27 \textpm {\scriptsize 0.32} & 98.10 \textpm {\scriptsize 0.45} & 98.10 \textpm {\scriptsize 0.45} \\
\bottomrule
\end{tabular}
\vspace{1ex}

\end{table}

Table~\ref{tab:localization} demonstrates that localization is a highly separable task. Most models—including PatchTST, ResNet18, and TimeSformer-1D—achieve perfect scores in the Easy and Medium tiers and retain near-perfect performance in the Hard tier. ViT, Transformer, and LSTM also show strong results (F1 $>$ 98\%). MLP consistently underperforms, particularly in the Hard tier (F1: 80.03\%), likely due to limited spatial modeling. Overall, most models handle localization with high reliability. These results indicate that CSI-based localization is a highly separable task, and that most temporal or spatially-aware models can solve it with high reliability.

Table~\ref{tab:motionsourcerecognition} shows consistently high motion source recognition performance across all difficulty levels. Most models achieve F1-scores above 96\%, with ViT and ResNet18 exceeding 99\% even in the Hard setting. MLP, PatchTST, and Transformer also perform well, indicating the task is relatively easy to separate. Performance variance remains low, suggesting stable generalization.

\subsection{Evaluation on OOD Splits}
\label{app:ood-results}

Tables \ref{tab:humanactivityrecognition}-\ref{tab:proximityrecognition} present the performance of supervised models under three cross-domain generalization settings—Cross-Device, Cross-Environment, and Cross-User—for Human Activity Recognition, Human Identification, and Proximity Recognition, respectively. Across all tasks, ViT consistently achieves the highest performance, with the best F1-scores in most OOD settings.

\begin{table}[t]
\centering
\caption{\textbf{Human Activity Recognition} cross-domain performance. Accuracy (Acc) and F1-score are reported as mean ± std (\%) over three runs.}
\label{tab:humanactivityrecognition}
\small
\begin{tabular}{l cc cc cc}
\toprule
\multirow{2}{*}{\textbf{Model}} & \multicolumn{2}{c}{\textbf{Cross-Device}} & \multicolumn{2}{c}{\textbf{Cross-Env}} & \multicolumn{2}{c}{\textbf{Cross-User}} \\
\cmidrule(lr){2-3}\cmidrule(lr){4-5}\cmidrule(lr){6-7}
&Acc & F1 & Acc & F1 & Acc & F1 \\
\midrule
LSTM~\cite{LSTM1997} & 60.57 \textpm {\scriptsize 2.12} & 57.04 \textpm {\scriptsize 2.32} & 53.65 \textpm {\scriptsize 0.89} & 46.22 \textpm {\scriptsize 0.72} & 53.33 \textpm {\scriptsize 2.11} & 45.70 \textpm {\scriptsize 2.01} \\
MLP~\cite{MLP} & 56.33 \textpm {\scriptsize 1.23} & 50.79 \textpm {\scriptsize 1.11} & 52.15 \textpm {\scriptsize 0.85} & 43.45 \textpm {\scriptsize 1.40} & 52.06 \textpm {\scriptsize 0.54} & 42.05 \textpm {\scriptsize 0.97} \\
PatchTST~\cite{PatchTST2023} & 61.61 \textpm {\scriptsize 1.81} & 58.05 \textpm {\scriptsize 1.54} & 56.85 \textpm {\scriptsize 0.63} & 49.55 \textpm {\scriptsize 0.47} & 56.44 \textpm {\scriptsize 1.47} & 49.25 \textpm {\scriptsize 1.33} \\
ResNet18~\cite{kaiming2016} & 66.21 \textpm {\scriptsize 1.96} & 63.57 \textpm {\scriptsize 1.90} & 57.98 \textpm {\scriptsize 0.87} & 50.90 \textpm {\scriptsize 0.96} & 59.24 \textpm {\scriptsize 1.47} & 52.07 \textpm {\scriptsize 1.53} \\
TimeSformer-1D~\cite{TimeSformer2021} & 60.24 \textpm {\scriptsize 1.00} & 55.70 \textpm {\scriptsize 1.20} & 54.65 \textpm {\scriptsize 0.93} & 46.63 \textpm {\scriptsize 0.79} & 54.95 \textpm {\scriptsize 0.84} & 45.74 \textpm {\scriptsize 0.79} \\
Transformer~\cite{attention2017} & 61.82 \textpm {\scriptsize 0.95} & 57.80 \textpm {\scriptsize 0.78} & 54.92 \textpm {\scriptsize 0.98} & 47.17 \textpm {\scriptsize 1.12} & 54.72 \textpm {\scriptsize 0.84} & 46.67 \textpm {\scriptsize 1.00} \\
ViT~\cite{ViT2021} & 66.33 \textpm {\scriptsize 1.73} & 63.65 \textpm {\scriptsize 1.69} & 58.87 \textpm {\scriptsize 1.12} & 51.86 \textpm {\scriptsize 1.31} & 59.00 \textpm {\scriptsize 1.36} & 51.48 \textpm {\scriptsize 1.26} \\
\bottomrule
\end{tabular}
\vspace{1ex}
\end{table}
\begin{table}[t]
\centering
\small
\caption{\textbf{Human Identification} cross-domain performance. Accuracy (Acc) and F1-score are reported as mean ± std (\%) over three runs.}
\label{tab:humanidentification}
\begin{tabular}{l cc}
\toprule
\multirow{2}{*}{\textbf{Model}} & \multicolumn{2}{c}{\textbf{Cross-Device}} \\
\cmidrule(lr){2-3}
&Acc & F1 \\
\midrule
LSTM~\cite{LSTM1997} & 59.25 \textpm {\scriptsize 1.69} & 59.32 \textpm {\scriptsize 1.72} \\
MLP~\cite{MLP} & 57.31 \textpm {\scriptsize 1.61} & 57.15 \textpm {\scriptsize 1.45} \\
PatchTST~\cite{PatchTST2023} & 60.45 \textpm {\scriptsize 1.07} & 60.56 \textpm {\scriptsize 1.17} \\
ResNet18~\cite{kaiming2016} & 68.07 \textpm {\scriptsize 1.93} & 68.21 \textpm {\scriptsize 1.97} \\
TimeSformer-1D~\cite{TimeSformer2021} & 60.84 \textpm {\scriptsize 0.81} & 61.00 \textpm {\scriptsize 0.79} \\
Transformer~\cite{attention2017} & 59.94 \textpm {\scriptsize 0.77} & 59.81 \textpm {\scriptsize 0.96} \\
ViT~\cite{ViT2021} & 69.37 \textpm {\scriptsize 1.53} & 69.55 \textpm {\scriptsize 1.61} \\
\bottomrule
\end{tabular}
\vspace{1ex}

\end{table}

\begin{table}[t]
\centering
\caption{ \textbf{Proximity Recognition} cross-domain performance. Accuracy (Acc) and F1-score are reported as mean ± std (\%) over three runs.}
\label{tab:proximityrecognition}
\small
\begin{tabular}{l cc cc cc}
\toprule
\multirow{2}{*}{\textbf{Model}} & \multicolumn{2}{c}{\textbf{Cross-Device}} & \multicolumn{2}{c}{\textbf{Cross-Env}} & \multicolumn{2}{c}{\textbf{Cross-User}} \\
\cmidrule(lr){2-3}\cmidrule(lr){4-5}\cmidrule(lr){6-7}
&Acc & F1 & Acc & F1 & Acc & F1 \\
\midrule
LSTM~\cite{LSTM1997} & 24.89 \textpm {\scriptsize 2.97} & 24.29 \textpm {\scriptsize 3.02} & 28.64 \textpm {\scriptsize 1.08} & 26.76 \textpm {\scriptsize 1.02} & 29.20 \textpm {\scriptsize 0.55} & 23.83 \textpm {\scriptsize 1.34} \\
MLP~\cite{MLP} & 28.73 \textpm {\scriptsize 0.89} & 27.31 \textpm {\scriptsize 0.84} & 25.76 \textpm {\scriptsize 0.77} & 20.86 \textpm {\scriptsize 1.41} & 26.19 \textpm {\scriptsize 0.57} & 17.32 \textpm {\scriptsize 0.74} \\
PatchTST~\cite{PatchTST2023} & 28.13 \textpm {\scriptsize 2.17} & 26.60 \textpm {\scriptsize 1.38} & 26.42 \textpm {\scriptsize 1.88} & 25.35 \textpm {\scriptsize 1.63} & 28.86 \textpm {\scriptsize 0.67} & 23.15 \textpm {\scriptsize 0.76} \\
ResNet18~\cite{kaiming2016} & 31.19 \textpm {\scriptsize 5.81} & 27.93 \textpm {\scriptsize 4.95} & 30.67 \textpm {\scriptsize 3.06} & 28.01 \textpm {\scriptsize 3.78} & 32.67 \textpm {\scriptsize 1.50} & 27.64 \textpm {\scriptsize 2.53} \\
TimeSformer-1D~\cite{TimeSformer2021} & 27.95 \textpm {\scriptsize 2.04} & 25.85 \textpm {\scriptsize 2.67} & 29.73 \textpm {\scriptsize 2.99} & 27.93 \textpm {\scriptsize 3.97} & 31.19 \textpm {\scriptsize 0.87} & 26.98 \textpm {\scriptsize 1.48} \\
Transformer~\cite{attention2017} & 30.68 \textpm {\scriptsize 3.11} & 28.76 \textpm {\scriptsize 3.51} & 29.67 \textpm {\scriptsize 1.63} & 27.12 \textpm {\scriptsize 1.79} & 30.26 \textpm {\scriptsize 1.94} & 25.97 \textpm {\scriptsize 2.26} \\
ViT~\cite{ViT2021} & 32.04 \textpm {\scriptsize 1.95} & 30.11 \textpm {\scriptsize 2.12} & 30.83 \textpm {\scriptsize 2.36} & 28.62 \textpm {\scriptsize 2.51} & 31.54 \textpm {\scriptsize 1.66} & 26.94 \textpm {\scriptsize 1.77} \\
\bottomrule
\end{tabular}
\vspace{1ex}
\\
\end{table}

For \textbf{Human Activity Recognition} (Table \ref{tab:humanactivityrecognition}), performance drops significantly under all OOD axes, particularly in the Cross-Environment and Cross-User settings, where even the top-performing models (ViT and ResNet18) show F1-scores below 53\%. This highlights the challenge of domain shifts in activity classification. 

In \textbf{Human Identification} results (Table \ref{tab:humanidentification}), ViT again leads with a 69.55\% F1 under Cross-Device, followed closely by ResNet18, suggesting strong person-specific feature learning. 

Lastly, \textbf{Proximity Recognition} (Table \ref{tab:proximityrecognition}) is the most challenging task, with all models performing poorly across OOD conditions. Even the best-performing ViT model achieves only around 30\% F1, and large variances are observed, indicating poor robustness and generalization. 

Overall, these results reveal that while certain models like ViT and ResNet18 show relative resilience, significant performance degradation remains under distribution shifts, underscoring the need for more robust domain generalization strategies in CSI-based sensing tasks.


\end{document}